# A parameterised equation of state, glass transition and jamming of the hard sphere system

Hongqin Liu

Integrated High Performance Computing,

Shared Services Canada, Montreal, Canada

A Gamma-distribution based potential energy landscape (PEL) theory has recently been proposed for supercooled liquids and glasses. This new PEL theory introduces a singularity term in the equation of state (EoS) suitable for representing the pressure of a glassy or jammed system. Using this framework, a parameterised EoS, $Z(\eta_J)$, is developed with the random-jammed-packing fraction, $\eta_J$, as an input. This EoS is capable of accurately calculating the compressibility (pressure) across the entire metastable and glassy region from $\eta_J \cong 0.62$ to 0.66, while seamlessly passing through the stable fluid region. Two special cases (paths) are examined in detail. The first path exhibits a singularity at the random close packing $\eta_J = \eta_{rcp} = 0.64$, traversing the metastable region explored by most simulations. Various thermodynamic properties calculated are compared to simulation data, showing excellent agreements. The second case addresses the first analytical EoS for the ideal glass transition in the hard sphere system. Finally, the transport properties of the hard sphere fluid are modeled using the Arrhenius law and the excess entropy scaling law. It is found that both laws fail (with slope changing) at $\eta = 0.555$, where the heat capacity peaks and the contributions of inherent structures and jamming effects begin to emerge.

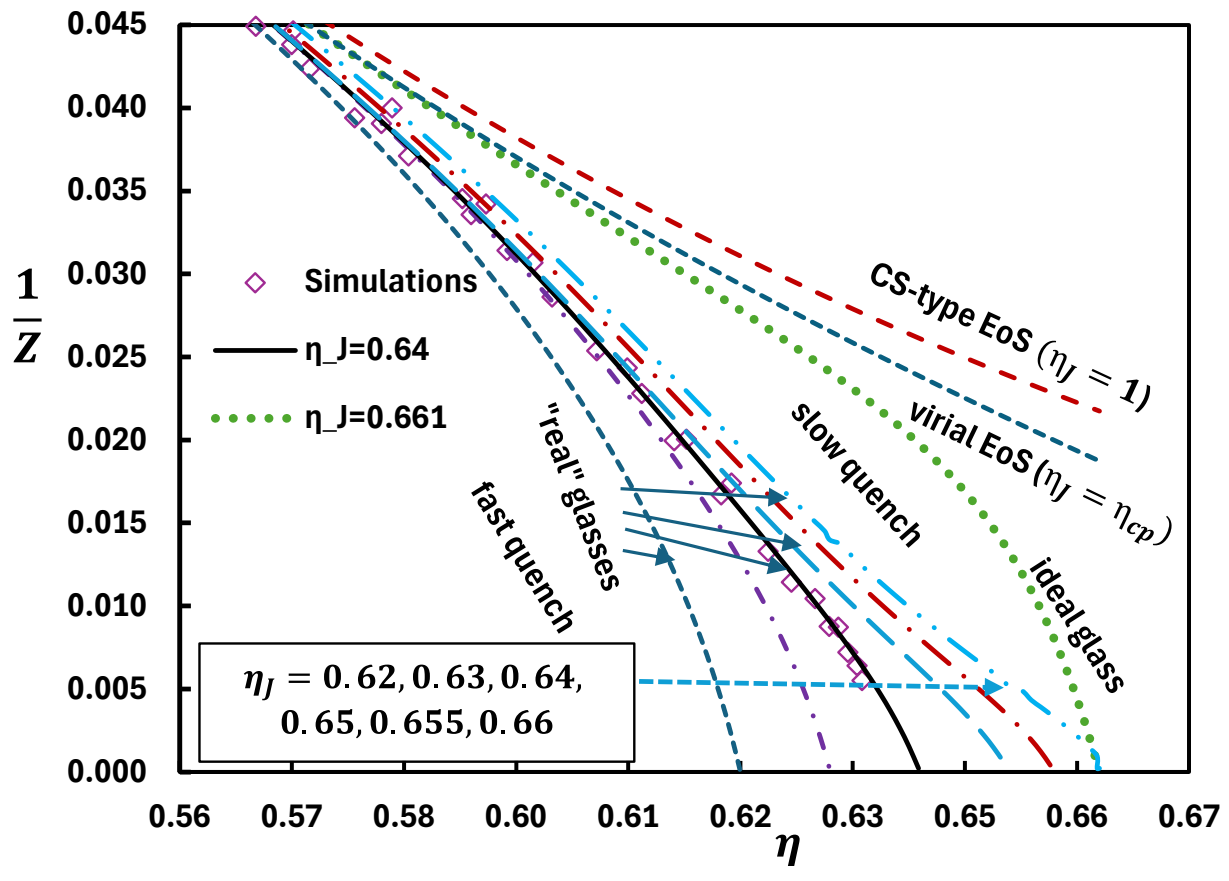

Emails: hongqin.liu@ssc-spc.gc.ca; hqliu2000@gmail.com.

## I. Introduction

Hard sphere (HS) glass transition has been an active research subject since late 1950s and early 1960s [1,2] and continues to attract significant attention [3-11]. Some comprehensive reviews on the topic are available in the literature [12-17]. The compression (or cooling) processes of a HSsystem that leads to glassy or jammed states is illustrated by Figure 1, where the compressibility factor $Z = P/(k_B T\rho)$, $P$ is the pressure, $T$, the temperature, $\rho = N/V$, the number ($N$) density, $V$, the total volume, $k_B$, the Boltzmann constant. The subscripts, "f", "m", "g", "rjp", "cp", refer to freezing, melting, glass transition, random jammed packing, close packing, respectively. The point $\rho^s$ indicates a density between the freezing density, $\rho_f$, and the glass transition density, $\rho_g$, from which the system is depressed via different paths up to different jammed states. Both $\rho^s$ and $\rho_g$ vary for different compression (quench) paths, but $\rho^{*s} (= \rho^s d^3) \approx 0.97$ ($d$ is the particle diameter) provides a close estimate for all paths.

The random close packing (rcp) refers to the densest random packing that lacks long-range order and may not necessarily exhibit mechanical stability. Typically, it takes the value (packing fraction) $\eta = \frac{\pi}{6}\rho d^3 = \frac{\pi}{6}\rho^* \approx 0.64$ [2,6,18]. In comparison, jammed packing is defined according to some mechanical stability criteria. It can be disordered or partially ordered, with the packing fraction values, $\eta$, from 0.62 up to 0.68 [6,8,9,14]. Here for convenience, we define the density range from the lowest random close packing to the highest as the random jammed parking (rjp) [15], as indicated in Figure 1. In the equations discussed below, we simply denote it as $\eta_J$ or $\rho_J^*$ [9].

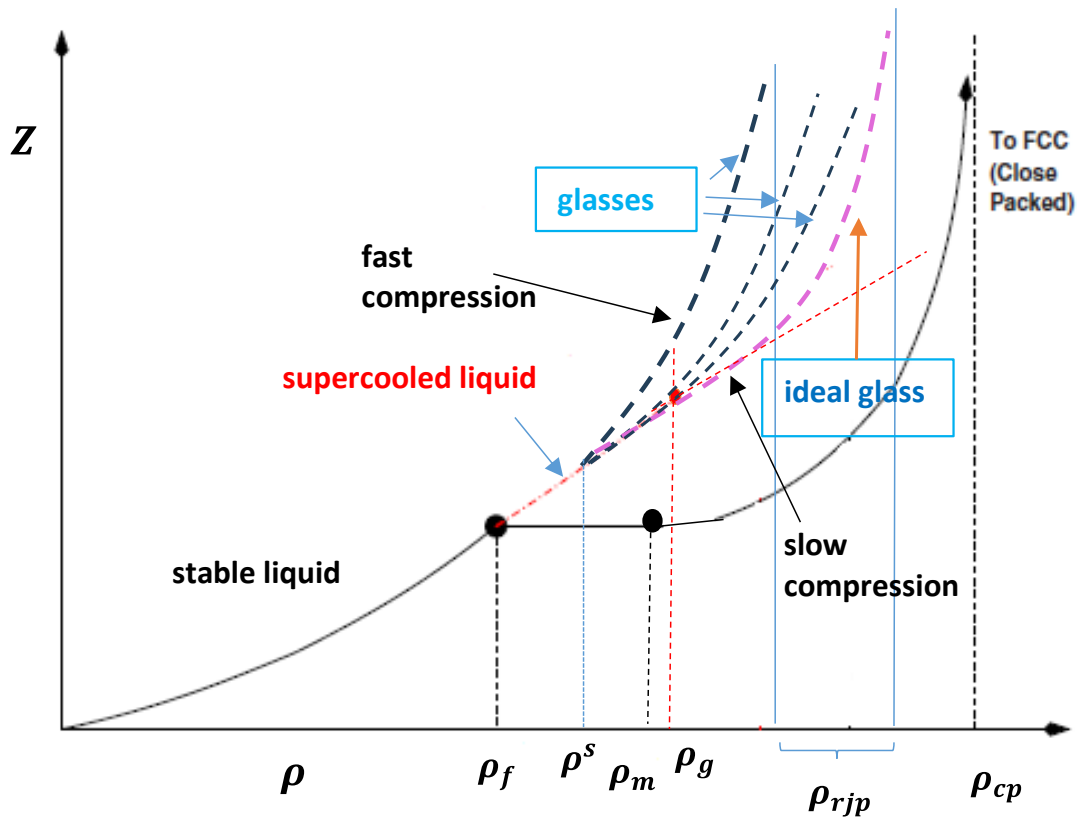

**Figure 1**. Phase diagram in the compressibility - density ($\rho$) space, see Refs.[14,15].

Speedy [4] illustrated the glass region based on his simulations and hypothetical equations of state (EoS) as shown by Fig.2.a. The projected heat capacity (Fig.2.b) for a glass is also depicted. Similar (to Fig.2.a) plots have also been adopted by other authors [8,15].

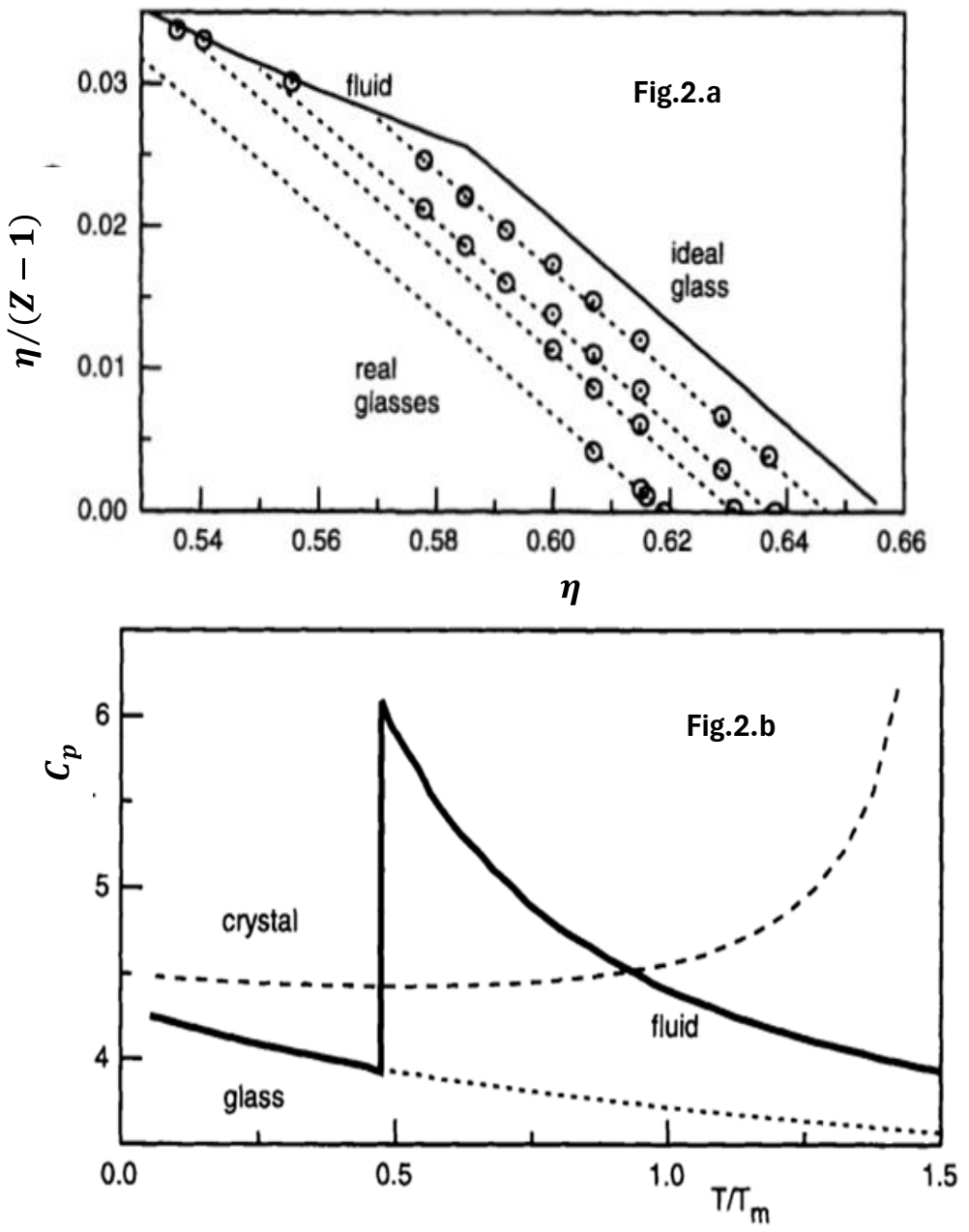

**Figure 2**. Projected EoS (Fgi.2.a) and heat capacity (Fig.2.b) by Speedy (1998), adapted from Ref.[4]. $T_m$ is the melting temperature, $C_P$, the heat capacity at constant pressure.

One of the most important findings from computer simulations is that the random jammed packing extends across a continuous region, from $\eta_J \sim 0.62$ to $\eta_J \sim 0.66$ [4], or even higher (0.68) [8,15]. In other words, there are infinite many paths from some point below the freezing point, $\rho^s$, to the random jammed packing (rjp) region. Assuming each path corresponds to a specific $\eta_J$ [4,9], there is a need for a "generic" or parameterised EoS that uses $\eta_J$ as an input, so that the entire area between the dotted lines in Figure 2.a can be produced. However, such a parameterised EoS is lacking and developing it constitutes the first objective of this work.

Secondly, computer simulations and theoretical analysis have revealed the existence of an ideal glass transition in the HS system [4,5,6]. However, an analytical EoS capable of predicting an ideal glass is still lacking. Addressing this gap is the second goal of the present work.

In addition, by utilizing a comprehensive database in the metastable (supercooled) fluid region, we will analyze the system behaviors, including pressure and other thermodynamic properties, along a specific path leading to the random close packing at $\eta_{rcp} = 0.64$. The majority of computer simulations and EoS studies have focused on this path. The extensive simulation data available will enable us to test and validate the EoS across a wide range of densities for various properties, including thermodynamic and transport properties.

To develop a physically robust EoS, we adopt the potential energy landscape (PEL) approach [21,22,23]. Currently, the Gaussian distribution is employed for configurational entropy [4,21-23], but it fails to predict the fragile-to-strong transition (crossover) [24] and results in EoS formulations that lack the necessary pole or singularity to account for pressure divergence. Such a singularity is a fundamental requirement for an EoS in a glassy system. In this work, we will propose a new EoS framework based on a Gamma-distribution based PEL theory [24].

The remainder of the paper is organized as follows: In the next section, we introduce the Gamma-distribution-based PEL theory and outline a framework for developing new equations of state (EoS). The third section presents a 'generic' EoS for the HS system, specifically propose a parameterized EoS that uses $\eta_J$ as an input to produce the entire 'glassy' region depicted in Figures 1 and 2, in accordance with the continuous $\eta_{rjp}$ region. In the fourth section, we present the calculation results and the database used for parameter estimation. Two special cases are discussed in detail: (a) a specific path leading to $\eta_J = 0.64$; and (b) the ideal glass transition. In the fifth section, we examine the glass transition in the context of transport properties. Lastly, the conclusions are summarized in the final section.

## II. Gamma-distribution based PEL theory and EoS

The potential energy landscape theory has been utilized to derive EoS for supercooled liquids and glasses [21,22,23]. The current PEL theories are mostly based on the Gaussian distribution for configurational entropy, which assumes a symmetric energy distribution. For metastable and glassy liquids, however, the distribution of the potential energy of inherent structures (IS) is inherently asymmetric [24]. Furthermore, the EoS derived from existing PEL theories lacks a singularity term [21-23], which limits their applicability to non-glassy states.

Figure 3 illustrates some key concepts and the relationships between the inherent structure (basin), glasses and the crystalline solid. While the system can explore numerous basins in configurational space, it becomes trapped at the glassy states (ideal or real), requiring extremely long time to "escape" and leading to pressure divergence. Similarly, at the stable crystalline state, the system remains permanently trapped and the pressure diverges as well. This framework highlights that an EoS for supercooled fluids and glassy systems should incorporate multiple components: (1) contributions from basins or IS; (2) contributions from glassy states; and (3) contributions from the crystalline solid.

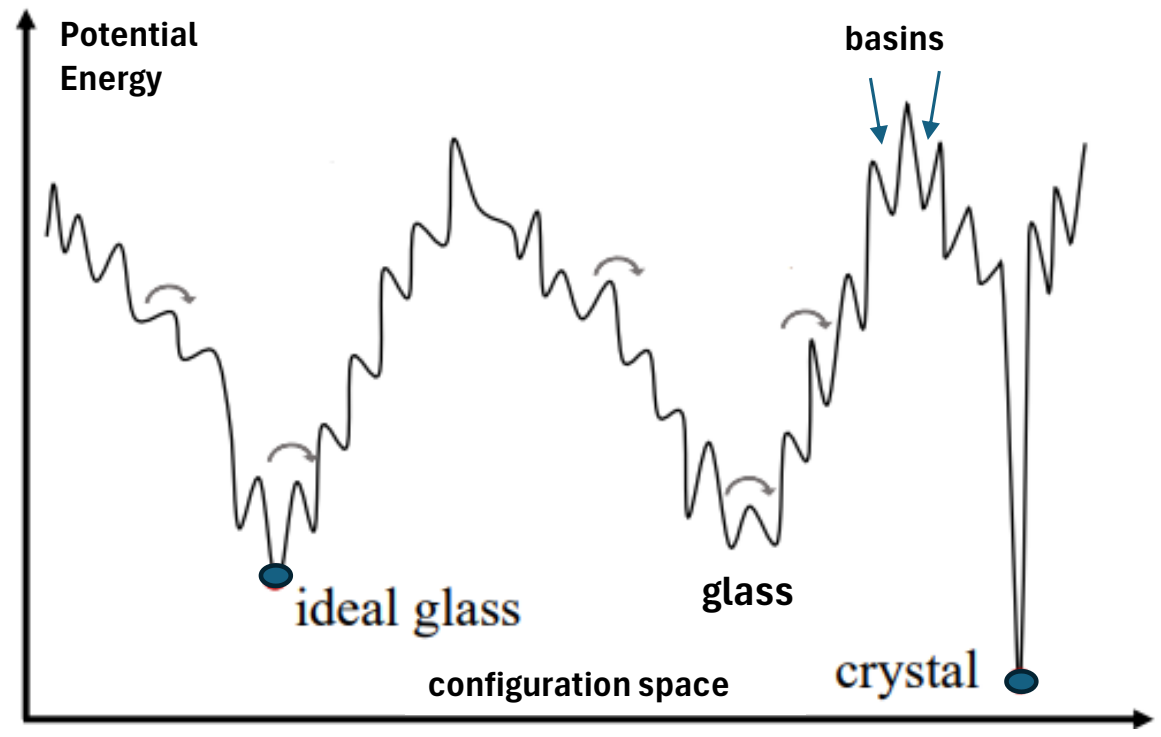

**Figure 3**. Illustration of potential energy landscape in configuration space

Within the PEL framework, the partition function of the system can be written as [25,26,27]:

$$Q(N,T,V) = \int exp\{-\beta[\epsilon_{is} - TS_c(N,V,\epsilon_{is}) + F_{vib}(N,T,V,\epsilon_{is})]\}d\epsilon_{is} \quad (1)$$

where $S_c$ is the configurational entropy, $F_{vib}$, the vibrational free energy, and $\epsilon_{is}$ is the potential energy (a local property) associated with a given inherent structure (basin). The key component of the theory is the configurational entropy. To represent the configurational entropy, $S_c$, one starts with the local potential energy distribution of the inherent structures ($\epsilon_{IS}$) at a given temperature and volume. Currently, the Gaussian distribution is most commonly adopted [25,26]:

$$\Omega(\epsilon_{is};\sigma,E_0,c_0) = \frac{e^{c_0 N}}{\sigma\sqrt{2\pi}} exp\left[-\frac{1}{2}\left(\frac{\epsilon_{is}-\epsilon_0}{\sigma}\right)^2\right] \quad (2)$$

where $\epsilon_0$ is the mean (expected) value of the energy, $\sigma$, the variance or the scaling parameter, the quantity, $e^{c_0 N}$, accounts for the total number of the inherent structures, $c_0$ being a constant depending on the volume. Then the configurational entropy is given by $S_c(N,V,\epsilon_{is})/k_B = ln\Omega(\epsilon_{is};\sigma,E_0,c_0)$. The local vibrational free energy $F_{vib}(N,T,V,\epsilon_{is})$ is usually assumed to be a linear function of the potential energy, $\epsilon_{is}$ [25,26]:

$$F_{vib}(N,T,V,\epsilon_{is}) = a_v + b_v\epsilon_{is} \quad (3)$$

where the parameters $a_v$ and $b_v$ are volume-dependent. In the PEL formalism, the system in equilibrium samples a narrow range of $\epsilon_{is}$, and one adopts the saddle point approximation at $\epsilon_{is} = E_{IS}$, which is determined from the lowest free energy constraint, or equivalently [27],

$$1 - T\left[\frac{\partial S_c(N,V,\epsilon_{is})}{\partial\epsilon_{is}}\right]_{N,V,T} + \left[\frac{\partial F_{vib}(N,V,\epsilon_{is})}{\partial\epsilon_{is}}\right]_{N,V,T} = 0 \quad (4)$$

Eq.(2), Eq.(3) and Eq.(4) lead to the temperature dependence of $E_{IS}$ and $S_c \sim E_{IS}$ relation for a given configurational entropy expression. Finally, with the saddle point approximation, Eq.(1) can be simplified as:

$$Q(N,T,V) \approx exp\{-\beta[E_{IS} - TS_c(N,V,E_{IS}) + F_{vib}(N,V,T,E_{IS})]\} \quad (5)$$

Hence the total free energy can be written as:

$$F(N,V,T,E_{IS}) = (1+b_v)E_{is} - TS_c(N,V,E_{IS}) + a_v \quad (6)$$

The final expression for $S_c(N,V,E_{is})$ is obtained with $\epsilon_{is}$ being replace with $E_{is}$ and $\epsilon_0$ with $E_0$. A simple $E_{IS} \sim T$ relation, $E_{IS} = E_0 - \sigma^2(1+b_v)/(2T)$, is also derived. From Eq.(6) and the volume dependences of the parameters, $c_0$, $\sigma$, $E_0$, $a_v$ and $b_v$ (obtained from IS simulations), one can derive an EoS for the supercooled liquids and glasses [21,22,23]:

$$Z_{lg} = Z_{IS} + Z_{vib} \quad (7)$$

The details of the expressions of $Z_{IS}$ and $Z_{vib}$ can be found from refs.[21-23]. An evident limitation of Eq. (7) is that both $Z_{IS}$ and $Z_{vib}$ remain finite across the entire density range as Eq.(1) is used. As mentioned earlier (see Figure 1), at a glassy or jammed state, the pressure ($Z_{lg}$) must diverge. Consequently, the EoS derived from the Gaussian-distribution-based PEL, as given in Eq. (7), fails to accurately describe the behavior at the random jammed packing (rjp) states.

As previously mentioned, the Gaussian distribution, as expressed in Eq. (1), assumes that the potential energy distribution of inherent structures is symmetric across the entire configurational space. However, such symmetry only applies under homogeneous conditions (stable liquids). As the configurational space extends into deep metastable and glassy regions, deviations from a normal distribution become significant, and the energy distribution becomes increasingly heterogeneous [24]. As a result, the IS potential energy distribution is expected to be asymmetric, making the Gaussian distribution unsuitable. The Gamma distribution serves as a more appropriate alternative for this purpose. Using PEL notations, the Gamma distribution can be expressed as:

$$\Omega(\epsilon_{is}) = \frac{e^{cN}\lambda}{\Gamma(\alpha)}\left(\frac{\epsilon_{is}-\epsilon_0}{\sigma_+}\right)^{\alpha-1} exp\left(-\frac{\epsilon_{is}-\epsilon_0}{\sigma_+}\right) \quad (8)$$

where $\alpha$ is the shape parameter; $1/\lambda$, the scalar parameter; the parameter $\sigma_+$ functions as the variance; $\Gamma(\alpha)$ is the Gamma function; $e^{cN}$, the total number of inherent structures. By using Eq.(3), Eq.(4) and Eq.(8), we have:

$$\frac{S_c}{k_B} = S_0 + (\alpha-1)ln\left(\frac{E_{IS}-E_0}{\sigma_+}\right) - \frac{E_{IS}-E_0}{\sigma_+} \quad (9)$$

where $S_0 = ln[e^{cN}\lambda/\Gamma(\alpha)]$ and the temperature dependence of $E_{IS}$ is given by:

$$E_{IS} = E_0 + \frac{aT}{b+T} \quad (10)$$

where $a = (\alpha-1)\sigma_1$ and $b = \sigma_1(1+b_v)$. The parameters of the Gamma-distribution based PEL formalism are $S_0$, $\alpha$, $E_0$, $\sigma_+$, $b_v$. An additional parameter, $\alpha$, is introduced in comparison with the Gaussian distribution. As $\alpha \to 1$, the distribution becomes exponential, and as $\alpha \to \infty$, the distribution becomes Gaussian-like. Finally, from Eq.(3), Eq.(6) and Eq.(9), we can rewrite the free energy as:

$$F(N,V,T;E_{IS}) = F_0 - k_BT(\alpha-1)ln\left(\frac{E_{IS}-E_0}{\sigma_+}\right) + g\frac{(E_{IS}-E_0)}{\sigma_+} \quad (11)$$

where two volume-dependent parameters are defined as $F_0 \equiv -k_BTS_0 + (1+b_v)E_0$, $g \equiv [k_BT + a_v + (1+b_v)\sigma_+]$. Finally, the equation of state can be derived:

$$Z = \frac{P}{\rho k_BT} = \beta\rho\frac{\partial F_0}{\partial\rho} - \rho ln\left(\frac{E_{IS}-E_0}{\sigma_+}\right)\frac{\partial\alpha}{\partial\rho} + \frac{\rho}{k_B(b+T)}\left[\frac{\partial ga}{\partial\rho} + a(\alpha-1)\frac{\partial b}{\partial\rho}\right] - \frac{ga\rho}{k_B(b+T)^2}\frac{\partial b}{\partial\rho} - k_B^{-1}a(\alpha-1)\rho\frac{\partial\alpha}{\partial\rho} \quad (12)$$

In Eq.12) we assume that $V$ is the molar volume (taking $N$ as the Avogadro constant) and the density $\rho = 1/V$. In the high density region, $\frac{\partial\alpha}{\partial\rho} > 0$ [24], by a mathematical relation, $\lim_{x\to 0^+} ln(x) \to -\infty$, a singularity is found in the EoS:

$$-\lim_{E_{IS}\to E_0} ln\left(\frac{E_{IS}-E_0}{\sigma_+}\right) \to \infty \quad (13)$$

Namely, as $E_{IS} \to E_0$ the system can be considered as "trapped" in a glassy or jammed state. For a crystalline solid, $E_{IS} - E_0$ is equivalent to $T \to 0$, Eq.(10). For supercooled liquid or glasses, it is equivalent to an effective temperature $T_{eff} \to 0$ [24,26]. Strictly speaking, $E_0$ should be treated differently for fragile and strong liquids [24]. However, for simplicity, we use the same notation, $Z_J$, for both cases. Following similar arguments presented in refs [21,22,23], we propose a new type of EOS for supercooled liquids and glasses:

$$Z_{lg} = Z_{IS} + Z_{vib} + Z_J \quad (14)$$

where $Z_{IS}$ is the from the IS contribution ($\alpha$, $\sigma_+$, $S_0$, $E_0$), $Z_{vib}$ from the vibrational contribution ($a_v$, $b_v$) and the component for the "jammed" state is defined as

$$Z_J = -\rho ln\left(\frac{E_{IS}-E_0}{\sigma_+}\right)\frac{\partial\alpha}{\partial\rho} \quad (15)$$

For expression of $Z_J$, a widely adopted EoS is $Z_J \propto (\rho_J - \rho)^{-1}$ for supercooled liquids and glasses [4,28]. Additionally, Eq.(12) includes terms involving $(b+T)^{-1}$ and $(b+T)^{-2}$. For strong fluids, $b < 0$ [24], therefore at $T = -b$, the pressure diverges, $Z \to \infty$. However, since the HS system is classified as fragile (see the fifth section), $b > 0$, the divergence of the $(b+T)$ terms is not a concern in this case.

### III. A "generic" equation of state and glass transition for the HS system

The goal of this section is to propose an EoS that applies across the entire density range, from an ideal gas to the cjp state. Starting from Eq.(14) and including the contribution from the stable fluid, we have:

$$Z = Z_v + Z_{lg} = Z_v + Z_{IS} + Z_{vib} + Z_J \qquad (16)$$

For the stable fluid region, the most adopted EoS is the Carnahan-Starling (CS)-type EoS. For instance, the current author proposed an accurate CS-type EoS [30]:

$$Z = \frac{1 + \eta + \eta^2 - \frac{8}{13}\eta^3 - \eta^4 + \frac{1}{2}\eta^5}{(1-\eta)^3} \qquad (17)$$

The accuracy of Eq.(17) is more than an order of magnitude higher than that from the original CS EoS [30]. However, due to the unphysical pole at $\eta = 1$, the CS-type EoS is considered unsuitable for describing the metastable fluid and glassy regions. In this work, we adopt the so-called closed virial EoS [3,31] for $Z_v$:

$$Z_v = 1 + Z_v' + \frac{b_0\eta}{1 - \eta/\eta_{cp}} \qquad (18)$$

where $b_0$ is a constant, the close packing fraction: $\eta_{cp} = 0.74048$. The term $Z_v'$ is composed of virial coefficients, $B_n$, distracting the coefficients from the last term of Eq.(18):

$$Z_v' = \sum_{n=2}^{11} \left(B_n - b_0\alpha_{cp}^{n-1}\right)\eta^{n-1} \qquad (19)$$

where $\alpha_{cp} = 1/\eta_{cp} = 1.350475$. In this work, we use the virial coefficients up to the 11th, $B_2 = 4$, $B_3 = 10$, and for the rest is calculated with a quadratic function [30]:

$$B_n = -1.8752 + 1.2872n + 0.944n^2, n \geq 4 \qquad (20)$$

Theoretically, one can obtain $Z_{IS}$ and $Z_{vib}$ from density-dependence of parameters, $S_0$, $E_0$, $\sigma_+$ etc. [21-23], which requires accurate data for $S_c$ and $E_{IS}$ over the entire range of density. Due to lack of the simulation data for the HS system, we have to determine the function for $Z_{IS}$, $Z_{vib}$ and $Z_J$ primely based on semi-empirical analysis. For $Z_J$ we adopt the Salsburg-Wood EoS coupled with a high-order power function[15,28,29]:

$$Z_J = \frac{b_1\eta^{n_1}}{1 - \eta/\eta_J} \qquad (21)$$

Considering that, at low and intermediate densities, the system is far from the jammed state, the power function, $b_1\eta^{n_1}$, ensures that $Z_J$ has a minimal impact on the virial (equilibrium) contribution, Eq.(16).

To derive analytical expressions for $Z_{IS}$ and $Z_{vib}$, Shell et al. (2003) [23] began with a repulsive potential energy: $u \propto \rho^{n/3}$ , where $n$ is a constant from a soft-sphere repulsive force of the type $r^{-n}$ ($r$ is the position variable). From there, they determined the expressions for the PEL parameters, such as: $E_0 \propto \rho^{n/3}$, $\alpha \propto \rho^{n/3}$ etc. (based on the Gaussian distribution). Finally the pressure expressions are obtained. The inherent structure contribution is a van der Waals type EoS [23]:

$$Z_{IS} = c_{vdw}\rho^{\frac{n}{3}+1} - a_{vdw}\rho^2 \qquad (22)$$

where $c_{vdw}$ and $a_{vdw}$ are constants. the $c_{vdw}$ term accounts for the repulsive contribution, while the last term, the attractive contribution. Similarly, for the vibrational contribution, $P_{vib} \propto \rho^{\frac{n}{3}+1}$, [23].

In this work, for $Z_{IS}$ and $Z_{vib}$, following the arguments and analysis [21,23], we propose high-order density functions, $Z_{IS} \propto \rho^m$ etc. For a HS system, only repulsive contribution is involved and, as a result, we have the following:

$$Z_{IS} + Z_{vib} + Z_J = \frac{b_1 \eta^{n_1}}{1 - \eta/\eta_J} + b_2 \eta^{n_2} + b_3 \eta^{n_3} \quad (23)$$

Here we treat $Z_{IS}$ and $Z_{vib}$ differently: $n_2 \neq n_3$, and all 3 constants, $n_1$, $n_2$ and $n_3$ are considered as parameters. Notably, for the HS system, the values of $n_2$ and $n_3$ will be greater than that of $n$ for a soft-sphere system ($n = 12$ for the Lennard-Jones system [23]). Finally, we arrive at the EoS applicable across the entire density range:

$$Z = 1 + Z_v' + \frac{b_0 \eta}{1 - \eta/\eta_{cp}} + \frac{b_1 \eta^{n_1}}{1 - \eta/\eta_J} + b_2 \eta^{n_2} + b_3 \eta^{n_3} \quad (24)$$

The values of the parameters, $b_0$, $b_1$, $b_2$, $b_3$, $n_1$, $n_2$, and $n_3$ will be determined from fitting the simulation data for the stable fluid and metastable fluid regions. The equation contains two poles: $\eta_{cp}$ corresponding to the crystal solid, $\eta_J$ to a glass (or an ideal glass). The $b_2$ and $b_3$ terms represent the contributions of basins and vibrations, in accordance with the physical framework illustrated in Figure 3. The formalism of the EoS, Eq.(24), is physically sound. However, since the parameters have to be obtained from fitting simulation data, the final EoS is semi-empirical.

Given an EoS, other thermodynamic properties can be easily calculated. For example, the isothermal compressibility, $k_T$, is calculated with the following:

$$k_T^* \equiv k_T \left(\frac{k_B T}{d^3}\right) = \left(\eta \frac{\partial Z}{\partial \eta} + Z\right)^{-1} \rho^{*-1} \quad (25)$$

where the reduced density, $\rho^* = \rho d^3$. A related quantity is the structure factor $S_0 = S(q = 0)$:

The heat capacity at constant pressure can also be calculated from the compressibility factor:

$$C_p^{ex} = Z^2 \left(Z + \frac{\eta \partial Z}{\partial \eta}\right)^{-1} \quad (26)$$

where $C_p^{ex} = C_p - C_p^{ig}$, the superscript "ig" refers to the ideal gas. Combining Eq.(25) and Eq.(26) leads to a simple relation between the isothermal compressibility and the heat capacity:

$$C_p^{ex} = Z^2 k_T^* \rho^* \quad (27)$$

Therefore, if an EoS can accurately predict $k_T^*$ and $Z$, the accuracy of the heat capacity prediction will be guaranteed. As a result, the analysis based on the heat capacity or the derivative of $Z$ will be reliable. The importance of such a test (on $Z'$ related properties) will be further addressed in the next section for the CS-type EoS [30] and a highly accurate (for $Z$) EoS, mKLM EoS [20]. On the other hand, the excess entropy can also be calculated as an integration quantity:

$$s^{ex} \equiv \frac{S - S^{id}}{N k_B} = -I \quad (28)$$

and finally the reduced chemical potential (Gibbs free energy) is given by:

$$\mu^* = \frac{\mu}{N k_B T} = Z - 1 + I \quad (29)$$

where

$$I \equiv \int_0^\eta \left(\frac{Z - 1}{\eta}\right) d\eta \quad (30)$$

The integrations for most terms in Eq.(24) are straightforward. For the $b_1$ term, it is given below:

$$I_1 = b_1 \int_0^\eta \frac{\eta^{n_1 - 1}}{1 - c\eta} d\eta = -\frac{b_1}{c^{n_1}} \left[\sum_{k=1}^{n_1 - 1} \frac{(c\eta)^k}{k} + ln(1 - c\eta)\right] \quad (31)$$

where $c = 1/\eta_J$. From Eq.(25) to Eq.(31), all thermodynamic properties can be predicted and the reliability of the predictions will be guaranteed if $s^{ex}$ and $\mu^*$ are accurately predicted.

## IV. EoS results and discussions

### IV.1. A parameterised EoS

As mentioned earlier, the values of the parameters, $b_0$, $b_1$, $b_2$, $b_3$, $n_1$, $n_2$, and $n_3$ are obtained from fitting the simulation data for the compressibility factor. A great amount of simulation data is available for the stable fluid and the metastable fluid regions. Following a detailed analysis, we adopt the data listed in Table 1 for the stable fluid region and Table 2 for the metastable fluid region, respectively. Along the path ending at $\eta_J = 0.64$, the data from both the stable and metastable regions are used to fit the EoS, Eq.(24), while for all other paths ($\eta_J = 0.62, 0.625, 0.63, 0.635, 0.645, 0.65, 0.655, 0.66$), the data only from the stable region are used. Once the parameter values are obtained for the selected paths, their density (packing fraction)-dependences are fitted using polynomial functions. It is found that the parameters, $b_0$, $n_1$, $n_2$, and $n_3$ can be treated as density-independent (Table 3). The remaining 3 parameters, $b_1$, $b_2$ and $b_3$, are fitted using simple polynomial functions (see the Supplementary Materials, SM, Figure S1):

$$b_i(\eta_J) = \sum_{k=0}^{3} c_{i,k}\,\eta_J^k, \qquad i = 1,2,3 \qquad (32)$$

Table 4 lists the correlation coefficients, $c_{i,k}$. During the parameter fitting process, the best accuracies in the stable region (Table 1) were sought at all densities from $\eta_J = 0.62$ to $0.66$, while for the path $\eta_J = 0.64$ the accuracy in the metastable region (Table 4) are also considered but with less weights. Using the "generic" EoS, Eq.(24) along with Eq.(32), Table 3 and Table 4, the absolute average deviation (AAD) produced is only 0.00065%, regardless of input $\eta_J$. The calculated values of $Z$ match the computer simulations (Table 1) up to 5+ digits (see Table S1 for selected densities). This level of precision is comparable to highly accurate simulations with 1 million particles [20,31]. For a comparison, the original Canahan-Starling EoS yields AAD=0.17%, while the improved CS-Type EoS, Eq.(18) [30] achieves an AAD of 0.0057% for the same data points listed in Table 1.

With the parameterised EoS, we can calculate the compressibility factor and other thermodynamic properties at any given random jammed packing, $\eta_J$. As a result, the entire area projected by Speedy (1998) [4] (Figure 2.a), and similarly by Jadrich and Schweizer (2013) [8], can be numerically produced across all densities from $\eta_J = 0.62$ to $0.66$. To the best of author's knowledge, this is the first EoS capable of numerically "painting" the entire metastable and glassy area.

Figure 4 shows a plot of $1/Z$ vs packing fraction in the high density region. Results from the Canahan-Starling (CS) type EoS [30] and virial EoS (using the first 3 terms of the r.h.s. of Eq.(24)) are also included for comparison. The so called "ideal glass" will be discussed in detail later. Figure 5 provides a closer look at the region with the packing fraction from $\eta = 0.57$ to $0.63$ in $Z \sim \eta$ space. Figure 6 plots the compressibility vs packing fraction in the $\eta = 0.51$ to $0.56$ region. From Figure 6, we see that all lines converge at density $\rho^* \approx 0.97$ ($\eta = 0.51$), and this is the point, $\rho^s$, highlighted in Figure 1. In the following, we discuss two special cases: (1) the path leading to the random close packing at $\eta_J = 0.64$; (2) an ideal glass along a special path to be determined.

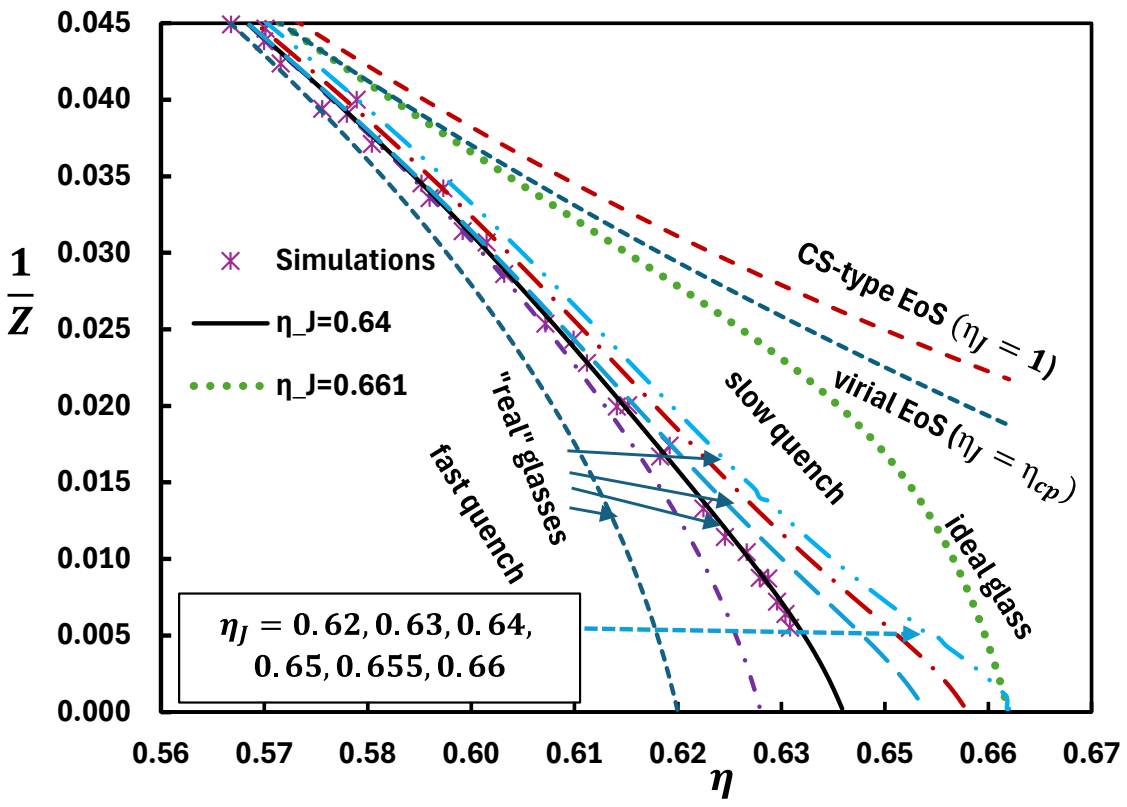

**Figure 4**. Calculation results for $1/Z$ at the selected packing fractions by Eq.(24) and Eq.(32). Data points are from simulations (Table 2).

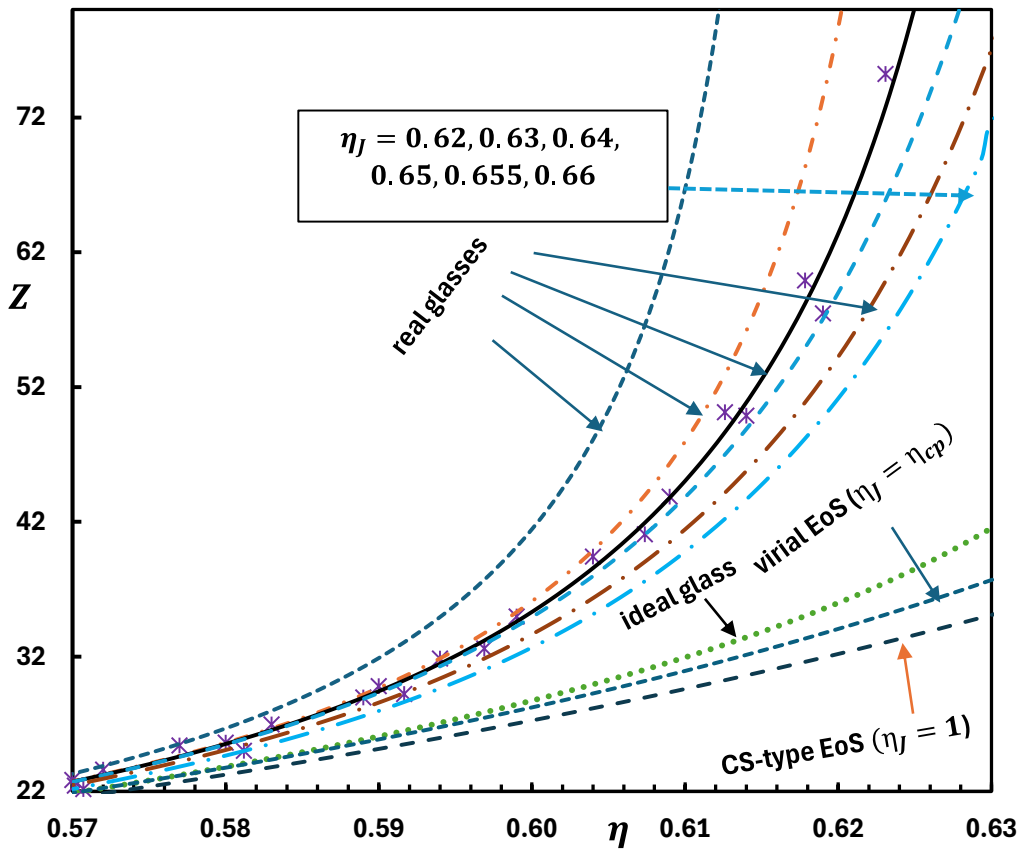

**Figure 5**. Plots of $Z \sim \eta$ relation in $\eta = 0.57$ to $0.63$ region.

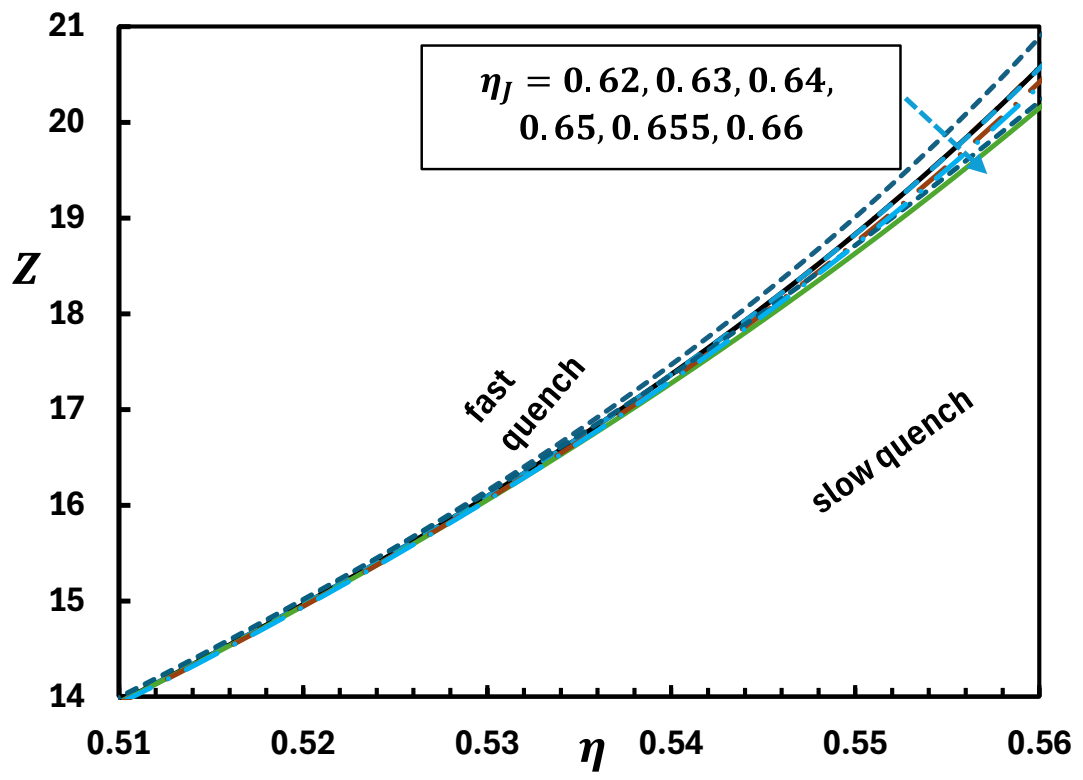

**Figure 6**. $Z \sim \eta$ plot for $\eta = 0.51$ to $0.56$ region. All curves merges at $\eta \approx 5.1$.

**Table 1** data sources for parameter fitting and testing in the stable region

| **data author** | NDP | rho* |
|---|---|---|
| Bannerman et al., 2010 [31] | 35 | 0.1-0.96 |
| Pieprzyk et al., 2019 [20] | 59 | 0.05-0.938 |
| Kolafa et al., 2004 [32] | 24 | 0.2-0.96 |

* NDP=number of data points.

**Table 2** data sources for the metastable region

| **authors** | NDP | rho* | role |
|---|---|---|---|
| Bannerman et al., 2010 [31] | 5 | 0.97-0.995 | fitting & testing |
| Pieprzyk et al., 2019 [20] | 41 | 0.94-1.02 | fitting & testing |
| Kolafa et al., 2004 [32] | 7 | 0.97-1.03 | fitting & testing |
| Rinto&Torquato, 1996 [33] | 7 | 0.95-1.13 | fitting & testing |
| Kolafa 2006 [19] | 8 | 1.02-1.09 | fitting & testing |
| summary | **68** | **0.94-1.13** | |
| Speedy 1997 [34] | 19 | 0.9-1.09 | testing |
| Woodcock, 1981 [3] | 15 | 0.94-1.05 | testing |
| Wu & Sadus, 2005 [35] | 28 | 0.97-1.21 | testing |
| Hermes&Dijkstra, 2010 [36] | 20 | 1.0-1.21 | testing |
| Zhou&Schweizer, 2019 [37] | 10 | 0.95-1.13 | testing |

**Table 3.** Contants in Eq.(24)

| $b_0$ | $n_1$ | $n_2$ | $n_3$ |
|---|---|---|---|
| 7.3 | 16 | 17 | 37 |

**Table 4**. Correlation coefficients in Eq.(32)

| $\boldsymbol{i}$ | $\boldsymbol{c_{1,i}}$ | $\boldsymbol{c_{2,i}}$ | $\boldsymbol{c_{3,i}}$ |
|---|---|---|---|
| 3 | 0 | 0 | 2.08899E+12 |
| 2 | 89040.0377 | -389264.9286 | -4.08846E+12 |
| 1 | -96627.5086 | 408044.7355 | 2.67216E+12 |
| 0 | 26819.9107 | -121443.6582 | -5.82537E+11 |

## IV.2. The path leading to $\eta_J = 0.64$

The EoS along this path is particularly important for two reasons. First, the majority of researchers have performed simulations of pressure (or compressibility) for the metastable HS fluid, and most studies have identified the random close packing as 0.64. Secondly, transport properties such as alpha relaxation time, viscosity and diffusion coefficients have also been simulated along this path [3,8,20,47]. We will briefly discuss the latter in the next section. For convenience, the parameter values are listed in Table 5 (and Table 3), where the values for $b_1$, $b_2$ and $b_3$ are calculated using Eq.(32).

**Table 5**. The EoS parameters, Eq.(24), for the path $\eta_J = 0.64$

| $b_1$ | $b_2$ | $b_3$ | $AAD\%$ |
|---|---|---|---|
| 1.44910E+03 | -1.97379E+04 | 6.28379E+08 | 0.00065 |

Note: $AAD\% = \frac{100}{NDP} \sum_{i=1}^{NDP} abs\left(\frac{Z_{cal} - Z_{sim}}{Z_{sim}}\right)$

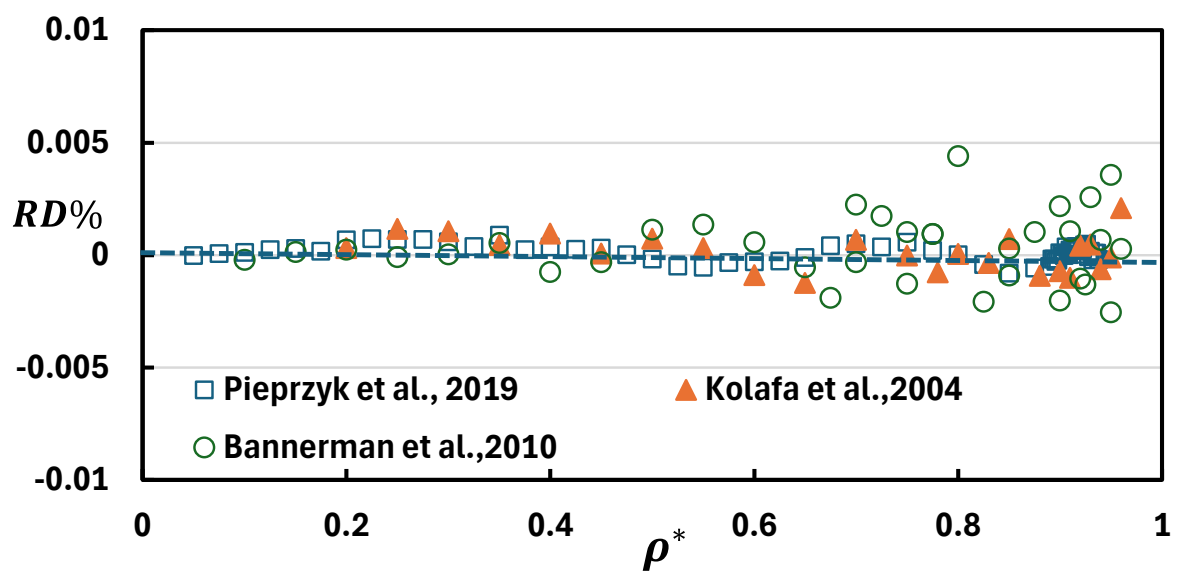

**Figure 7**. The relative deviations for the stable fluid region. $RD\% = 100 * (Z_{cal} - Z_{sim})/Z_{sim}$.

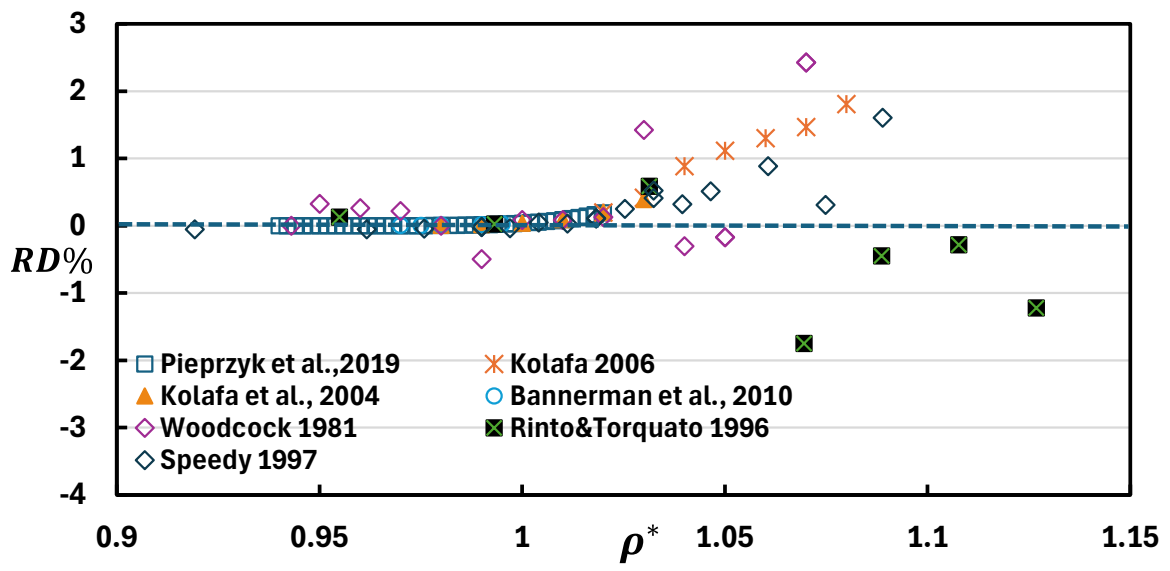

**Figure 8**. The relative deviations for the metastable fluid region.

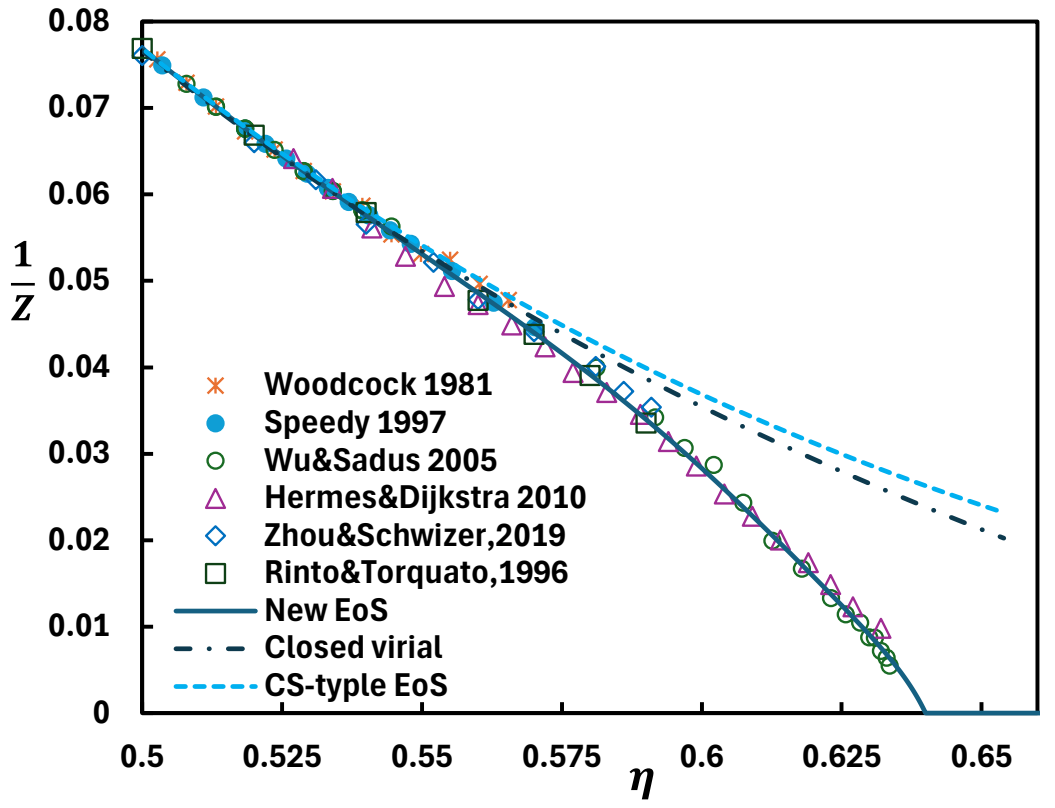

**Figure 9**. plots of $1/Z \sim \eta$ in the metastable region, calculated vs simulation data.

Figure 7 depicts the relative deviations (RD%) from Eq.(24) for the three data sets employed in this work (Table 1). We see that the different data sets are highly consistent with each other. The RD% values are less than 0.001% up to $\rho^* \sim 0.7$, and it start to increase as $\rho^* \sim 0.85$. Figure 8 shows the RD values for the metastable region (Table 2).

Figure 9 compares the calculated results and simulation data in terms of $1/Z \sim \eta$. As $\rho^* \sim 1.0$, the data from various sources are consistent with each other and as $\rho^* > 1.02$, data become increasingly scattered. Despite this, the agreement between data from different authors remains acceptable (see SI, Table S5).

To further test the new EoS, Eq.(24), prediction results for excess entropy, chemical potential, and isothermal compressibility are presented in Figure 10, Figure 11 and Figure 12, respectively. The excess entropy and isothermal compressibility are extended into the deep metastable regions. As shown the figures, the agreements with simulation data are excellent. These highly accurate predictions along with the exceptional accuracy of compressibility factor in the stable fluid region (0.00065% AAD) demonstrate the reliability of the EoS and the analysis based on it.

Figure 13 depicts the first and second derivatives of $Z$ and no abnormalities are observed. Figure 14 plots the heat capacity at constant pressure, calculated from Eq.(26). The heat capacity peaks at $\eta = 0.55$. Meanwhile, calculations and simulation results for compressibility factor, entropy, and isothermal compressibility reveal no indication of any 'phase' transition in the metastable region, including during the stable-to-metastable crossover. The only notable feature is the peak observed in the heat capacity.

As a result, we conjecture that the glass transition is a high-order transition, beyond the 2nd order. By the way, from Figure 14 and Eq.(33) (below), we can compute the configurational entropy along this path: $S_c = 0.222$. Since the freezing entropy $\Delta S_f = 1.168$ [45], the residue entropy stored in the glass is $\Delta S_r \approx 0.95$, which is comparable to the value estimated by Finney and Woodcock (2014): $\Delta S_r = 1.0$ [44], where the random close packing is defined as $\eta_{rcp} = 0.6366$.

One of the most accurate equations of state (EoS) for HS fluids, covering both the stable and metastable regions, is the one proposed by Kolafa et al. (2004) [32], known as the KLM EoS. Pieprzyk et al. (2019) [20] modified this EoS and proposed the mKLM EoS. This is a highly accurate EoS, achieving an AAD of 0.00063% in the stable region for the data sets listed in Table 1, matching the accuracy of Eq. (24).

However, the use of a polynomial function of $\eta/(1-\eta)$ in the EoS introduces a significant drawback: the compressibility reaches a maximum ($Z = 24.47$) at $\rho^* = 1.087$ and eventually drops to $Z = 0$ at $\rho^* \approx 1.1369$ (see Figure S2). This numerical issue leads the observation that the second derivative, $Z''_\eta$, exhibits a maximum [20]. As a result, the mKLM (and KLM) EoS is not applicable for $\rho^* \geq 1.09$ ($\eta = 0.57$) and any physical analysis [20] based on the behavior of $Z'_\eta$ and $Z''_\eta$ from the EoS may not be reliable in that region.

In computer simulations, as the density reaches the freezing point, the crystalline structures begins to form and the nucleation occurs. In spite of that, as demonstrated by many researches, the nucleation process can be avoided in slow or very slow quench processes allowing simulations to extend into the deep metastable region [8,33,34,35,36,37]. Consequently, in this work, we adopt a continuous and 'well-behaved' EoS that extends from the ideal gas state to random close packing (jammed), as suggested by many authors [8,9,12,44]. Here, the term 'well-behaved' refers to an EoS with no sign changes in its derivatives, (up to $Z'''_\eta$), throughout the entire region.

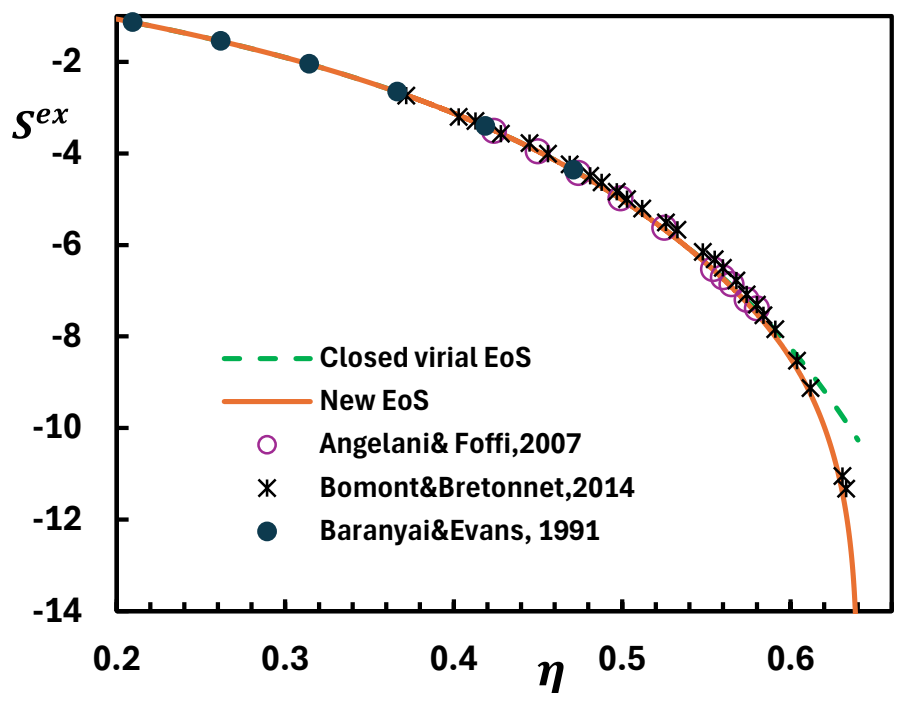

**Figure 10**. Prediction of the excess entropy, Eq.(24) and Eq.(28), vs simulation data, Ref.[38,39,40].

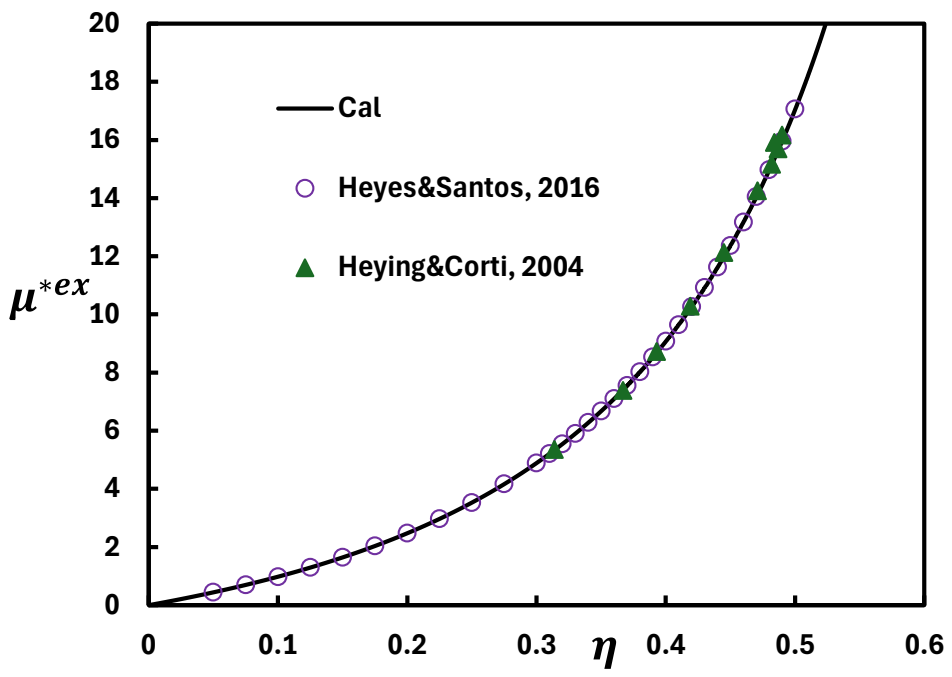

**Figure 11**. Prediction of the excess chemical potential (Gibbs free energy), Eq.(29), vs simulation data [41,42].

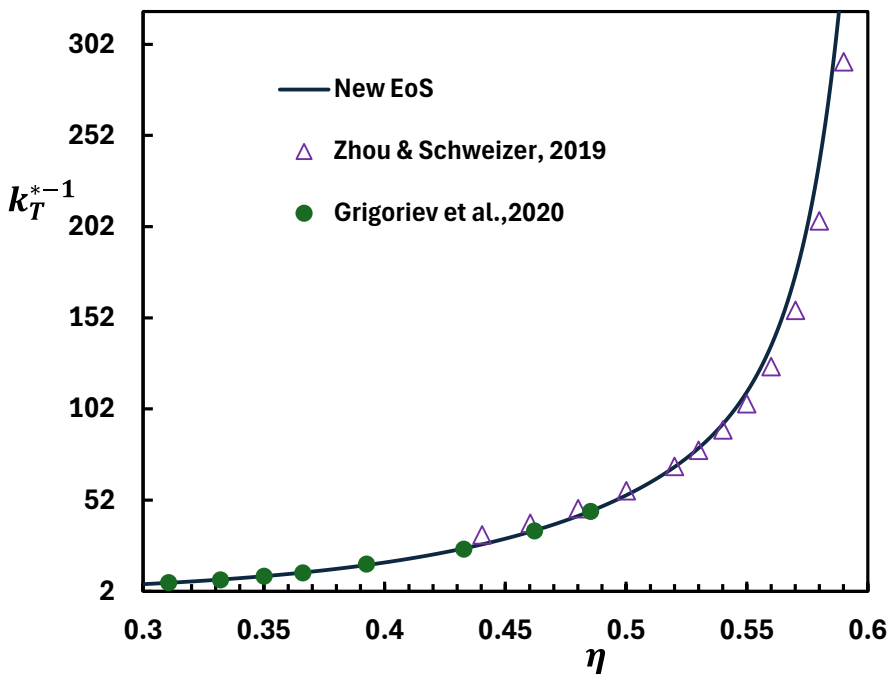

**Figure 12**. Prediction of the isothermal compressibility. Solid line from Eq.(24), Eq.(25). Data from Ref. [37,43].

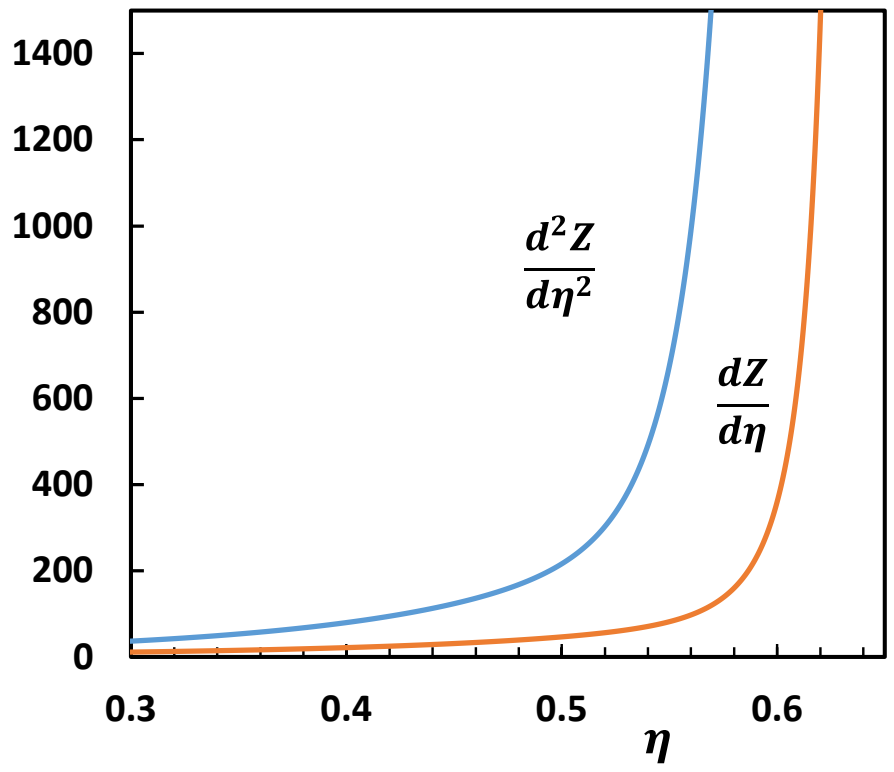

**Figure 13**. The first and second derivatives of compressibility by Eq.(24) and Table 5.

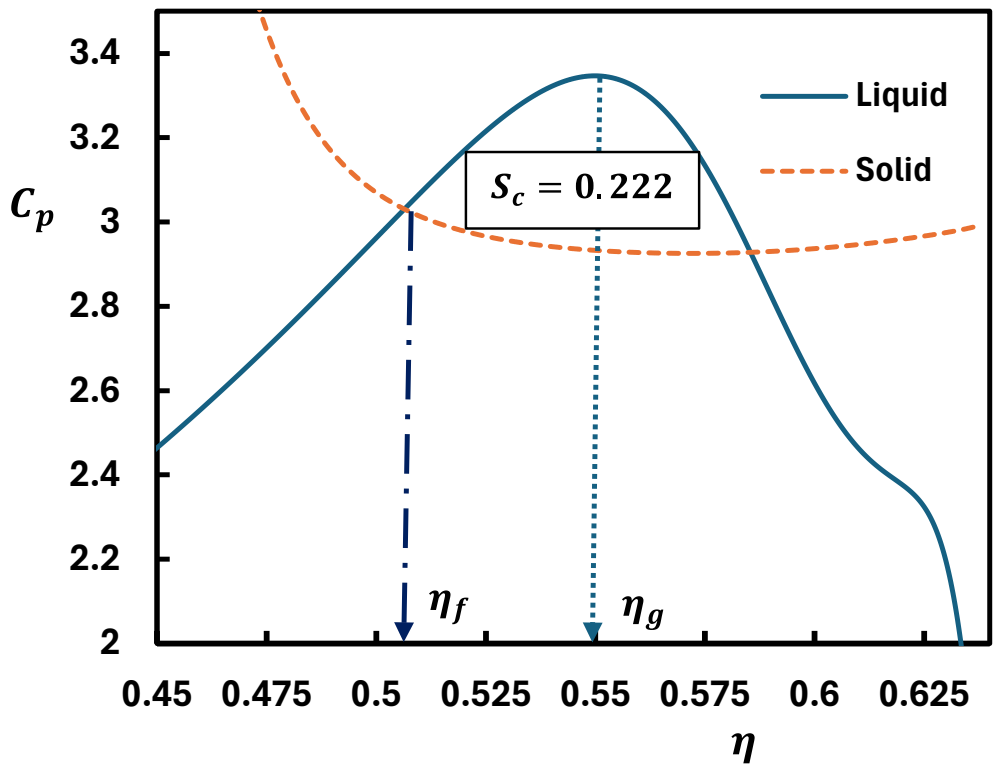

**Figure 14**. Heat capacity at constant pressure and the glass transition. The solid line is calculated by Eq.(27) with Eq.(24) for the liquid phase; the dashed line is calculated by Eq.(26) for solid phase where the EoS is from Ref.[31] (see SM).

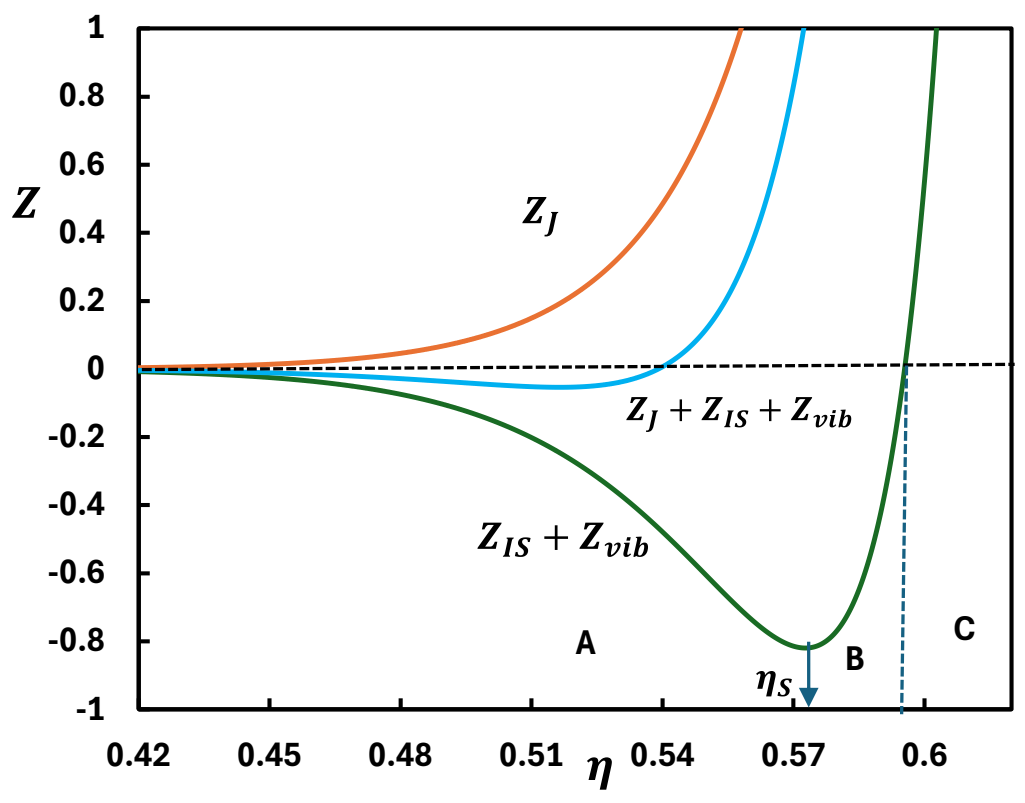

**Figure 15**. The compressibility factor of inherent structures and the Sastry point.

Finally, we discuss an intriguing feature of the inherent structures in terms of the compressibility calculated by the new EoS, Eq.(24). Figure 15 plots the contributions from $Z_{IS} + Z_{vib}$ and $Z_J$. An observation is that as $\eta \leq 0.42$ $(\rho^* = 0.8)$, contributions from all three components diminishes leaving the system pressure to originate solely from the stable (virial) fluid contribution. In other words, the inherent structures begin to form as density $\rho^* > 0.8$.

In a system characterized by both repulsive and attractive interactions, such as the van der Waals type, Eq.(22) [23]. This structure of the EoS exhibits a minimum point at a density, known as the Sastry density, $\rho_S$, which is the density of limiting mechanical stability or the point of ultimate strength [23]. As $\rho > \rho_S$, the system becomes glassy and mechanically less stable. For the HS fluid, there is no attractive interaction, hence $Z_{IS} > 0$ across the entire density region.

Nevertheless, Figure 15 shows that there is a minimum point (negative) on the $Z_{IS} + Z_{vib}$ curve at $\eta = 0.573$, which can be considered as the Sastry point for the HS system. In other words, at $\eta = 0.573$ the system becomes most mechanically stable.

Another point of interesting is at $\eta \approx 0.55$ $(\rho^* = 1.05)$ where the sum $Z_J + Z_{IS} + Z_{vib}$ transitions from negative to positive. This marks the point at which the CS-type EoS and the virial EoS begin to fail (see Table S5), and the heat capacity peaks. In summary, from the ideal gas to the freezing point, $\eta_f = 0.492$ $(\rho^* = 0.94)$, the system exists in stable (equilibrium) states; from $\eta = 0.492^+$ to 0.55, the system transitions into metastable (supercooled) states; as $\eta > 0.55$ the system enters the glassy states, suggesting the glass transition point: $\eta_g \approx 0.55$. Finally, at $\eta = 0.64$ the system becomes jammed.

### IV.3. The ideal glass transition

The ideal glass transition in the HS system has been discussed in the literature from different perspectives [4,6,23]. However, an analytical EoS capable of predicting the ideal glass transition remains unavailable. During the compression process, an ideal glass is formed under two constraints: (1) the system stays in stable states as much as possible; (2) the freezing entropy is fully converted into the configurational entropy at a specific point, known as the Kauzmann point. The configurational entropy can be calculated from heat capacity [3] as follows:

$$S_c(T) = \Delta S_c = \int_0^{T^*} \frac{C_p(fluid) - C_p(crystal)}{T^*} dT^* \quad (33)$$

For the HS system, the "temperature" is defined as [3] $T^* = 1/(\rho Z)$. Similar to Eq.(33), one can calculate the Kauzmann temperature, $T_K^*$:

$$\Delta S_f = \int_{T_K^*}^{T_f^*} \frac{C_p(fluid) - C_p(crystal)}{T^*} dT^* \qquad (34)$$

where $T_f^*$ is the freezing temperature, $\Delta S_f$, the melting entropy. The integrations in Eq.(33) and Eq.(34) can be evaluated using numerical methods. The heat capacity is calculated by Eq.(26). For the liquid branch Eq.(24) is employed, and for the solid phase the EoS proposed in Ref.[20] is utilized. The calculation details are provided in the Supplementary Material. The melting entropy for the HS fluid is taken from a relative recent simulation [45]:

$$\Delta S_f = 1.168 \qquad (35)$$

The parameterised EoS discussed above is not applicable to the ideal glass, as the constraint, Eq.(34), was not included. To develop an EoS for the ideal glass, all parameters are re-fitted using a target objective function composed of two components: (a) deviations for the data points listed in Table 1 (the stable region); (b) the difference between Eq.(35) and Eq.(34). Table 6 lists the fitting results. As shown in the table, for the ideal glass, $\eta_J = 0.661$, and the AAD for the stable fluid region is 0.0015%. While this is little higher than that produced by the "generic" EoS, Eq.(24) and Eq.(32), it remains sufficiently accurate for our purposes.

**Table 6**. Values of the parameters of Eq.(24) for the path leading to the ideal glass

| $\eta_{igj}$ | $b_0$ | $n_1$ | $n_2$ | $n_3$ | $b_1$ | $b_2$ | $b_3$ | $AAD\%$ |
|---|---|---|---|---|---|---|---|---|
| 0.661 | 10.0 | 14 | 14 | 42 | 261.7775 | -3.170158E+03 | -6.52744E+08 | 0.0015 |

The compressibility factor along the ideal glass path is plotted in Figures 4 and 5. The figures show that the ideal-glass compression path closely follows the equilibrium path (virial EoS) up to $\eta \approx 0.6$, satisfying the constraint (1). There is a significant gap between the ideal path and the "regular" glass path ($\eta_J = 0.66$), which is attributed to the entropy constraint, Eq.(34).

Figure 16 illustrates the heat capacities of the ideal glass vs that of the crystalline solid. The heat capacity peaks at $\eta \approx 0.61$, which suggests that the ideal glass transition point is $\eta_g \approx 0.61$. Additionally, the heat capacities generated from the virial EoS (equilibrium) and a CS-type EoS [30] are also shown. The heat capacity along the ideal glass path closely matches that from the virial EoS (equilibrium) up to a deep metastable region, $\eta = 0.553$ ( $\rho^* = 1.0553$). In contrast, the heat capacity from the CS-type EoS begins to deviate from the equilibrium values at a density slightly above the freezing point, $\rho_f^* = 0.9392$ [20]. This suggests that even though the CS-type EoS can be used to estimate the pressure beyond the freezing point, up to $\rho^* = 1.05$, the derivative properties from the EoS are reliable only in the stable fluid region.

Finally, Figure 17 depicts the configurational entropy predicted by Eq.(33) compared with the simulation data [38] for the HS ideal glass. The agreement is satisfactory, demonstrating the reliability of the proposed EoS for the ideal glass path. From Figure 16 or Figure 17, we identify the Kauzmann temperature, $T_K^* = (\rho^* Z)^{-1} = 0.016$, where $C_p^{liquid}(T_K^*) = C_p^{solid}(T_K^*)$ or equivalently, packing fraction, $\eta_K \approx 0.64$, where $S_c(\eta_K) = 0$. The so-called Kauzmann crisis is evident: as all freezing entropy is depleted at $\eta_K$, the configurational entropy would become negative as $\eta > \eta_K$.

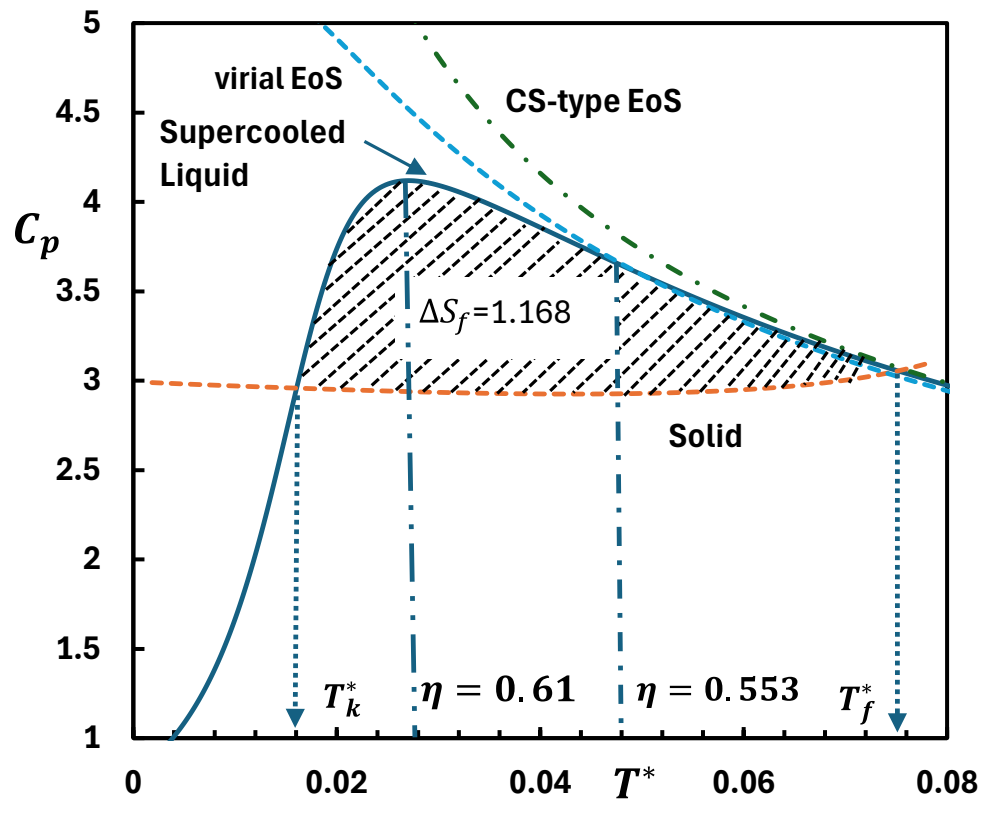

**Figure 16**. Heat capacity for the ideal glass path.

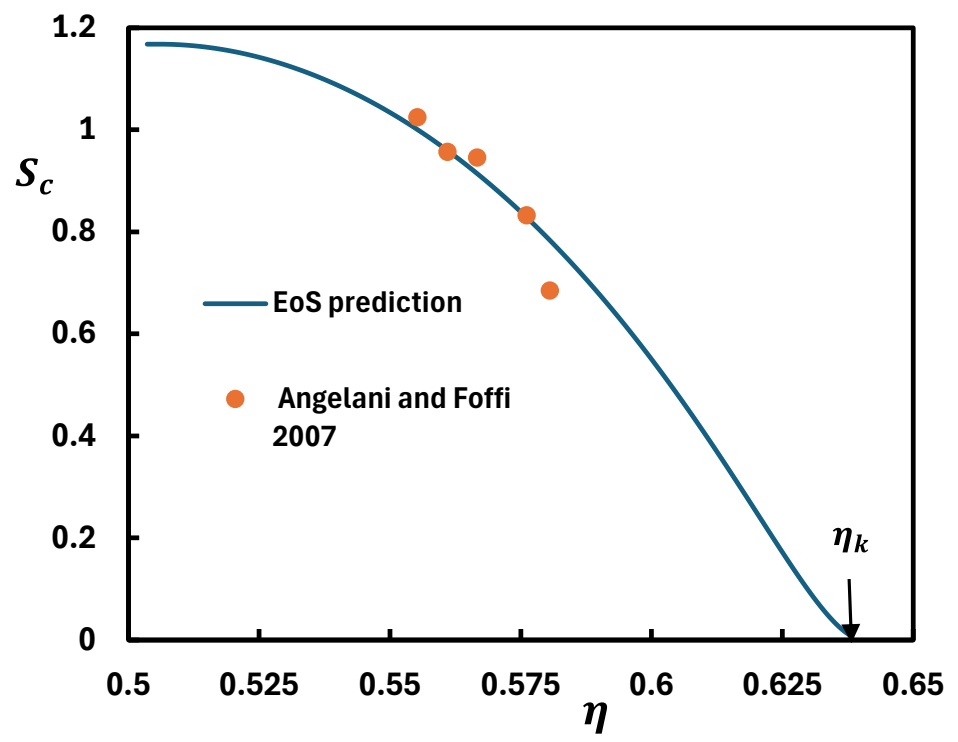

**Figure 17**. Configurational entropy of the ideal glass and the Kauzmann point: $\eta_K = 0.64$. Simulation data from ref.[38].

## V. Transport properties

In studies of glass transition, transport properties such as viscosity and diffusivity are frequently used. The concept of fragility has been introduced to describe the non-Arrhenius behaviors [46]. For the HS system, extensive simulations of the transport properties have been reported [3,8,20,47]. With the availability of an EoS, Eq.(24), encompassing the entire density range, we can now evaluate certain theoretical models for transport properties. Due to the complexity of the topic, here we only briefly explore the issue using the Arrhenius law and a free volume power law, or equivalently the entropy scaling law [48,49]. First we define a few quantities requires for the modeling. The viscosity of a dilute gas ($\mu_0$) is given by the following:

$$\mu_0 = \frac{1}{\emptyset_0} = \frac{5}{16d^2}\left(\frac{mk_BT}{\pi}\right)^{\frac{1}{2}} = \frac{5}{16\sqrt{\pi}}\frac{\sqrt{m\varepsilon\tilde{T}}}{d^2} \quad (36)$$

where $\emptyset_0$ is the fluidity of dilute gas, $d$, the particle diameter, $m$, the mass of the particle, $\varepsilon$, the energy parameter, $\tilde{T} = k_BT/\varepsilon$. The reduced viscosity or fluidity is defined as:

$$\mu^* = \frac{1}{\phi^*} = \frac{\mu}{\mu_0} = \frac{\emptyset_0}{\phi} \quad (37)$$

Similarly, for the $\alpha$ relaxation time, $\tau^* = \tau/\tau_0$ [8]. The diffusivity of dilute gas is given by

$$D_0 = \frac{3}{8\rho d^2}\left(\frac{kT}{\pi m}\right)^{\frac{1}{2}} = \frac{3}{8\pi^{1/2}\rho^*}\left(\frac{\tilde{T}d\epsilon}{m}\right)^{\frac{1}{2}} \quad (38)$$

The reduced diffusivities are defined as:

$$D^* = \frac{D}{D_0}, \qquad D^+ = \frac{D}{D_0 g(d)} \quad (39)$$

where $g(d)$ is the radial distribution function at contact ($r = d$), calculated from an EoS:

$$g(d) = \frac{Z-1}{4\eta} \quad (40)$$

Using the reduced diffusivity $D^+$ was proposed by Dzugutov [50] and turned out to be effective [48]. For the HS fluid, the Arrhenius plot for the viscosity (or the $\alpha$ relaxation time, $\tau/\tau_0$) can be written as:

$$ln\mu^* = -ln\phi^* = A_\mu + \frac{B_\mu}{T^*} = A_\mu + B_\mu(\rho^* Z) \quad (41)$$

where $A_\mu$ and $B_\mu$ are constants, and the relation $T^* = (\rho^* Z)^{-1}$ is valid for the HS system only. An interesting point is that Eq.(41) implies a linear relation, $ln\mu^* \propto (\rho^* Z)$. In comparison, a relation $ln\mu^* \propto (Z)$ has been proposed for more realistic fluids [7]. Similarly, for the diffusion coefficient:

$$lnD^* = A_d + \frac{B_d}{T^*} = A_d + B_d(\rho^* Z) \quad (42)$$

Or using the second reduced diffusivity, Eq.(39), we have:

$$lnD^+ = A_d^+ + \frac{B_d^+}{T^*} = A_d^+ + B_d^+(\rho^* Z) \qquad (43)$$

where $A_d$ and $B_d$ etc. are constants. In the following calculations, the compressibility factor $Z$ is computed from Eq.(24) with $\eta_J = 0.64$.

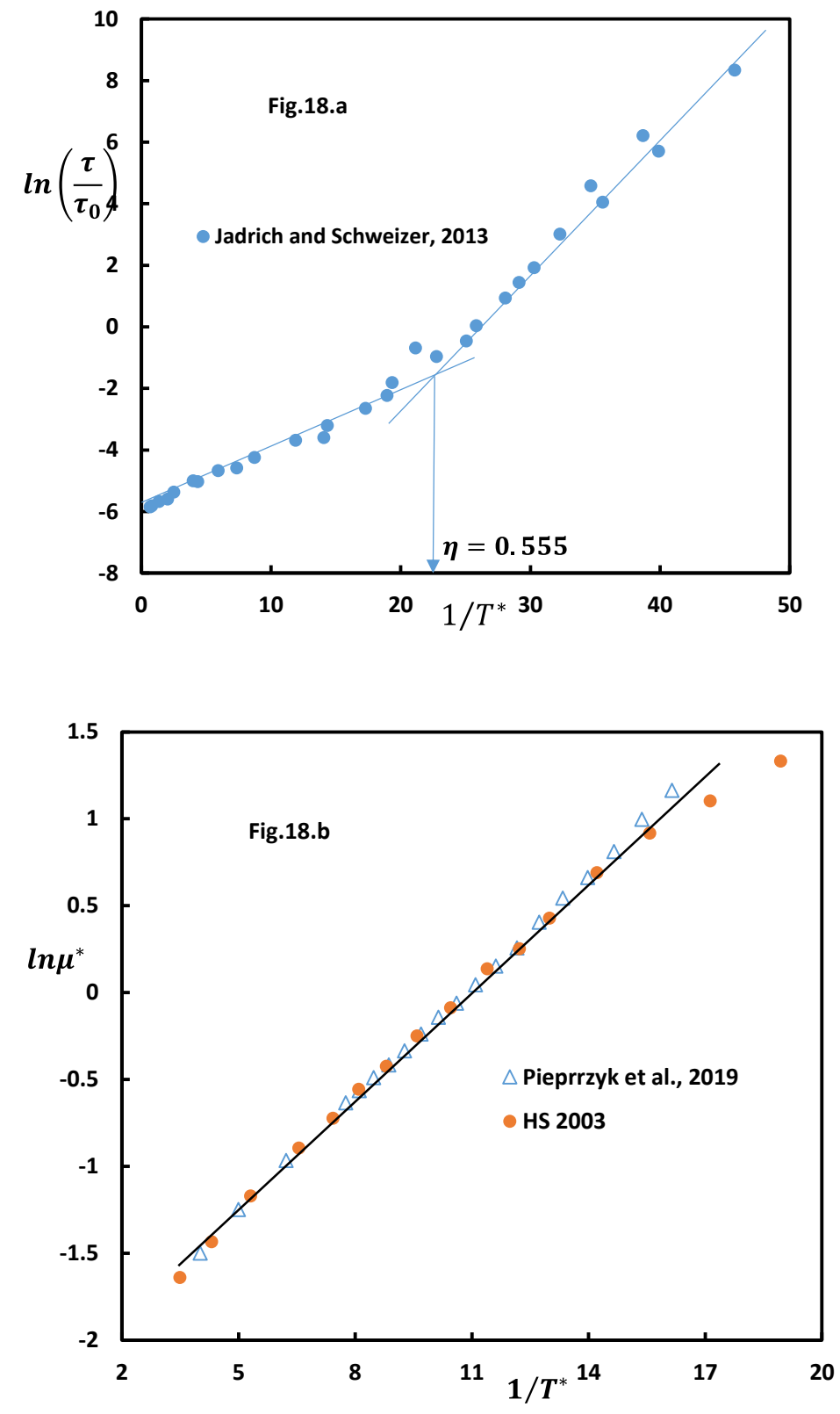

**Figure 18**. Arrhenius plots for $\alpha$ relaxation time (a) and viscosity (b), the data points are from Refs.[8,20,47].

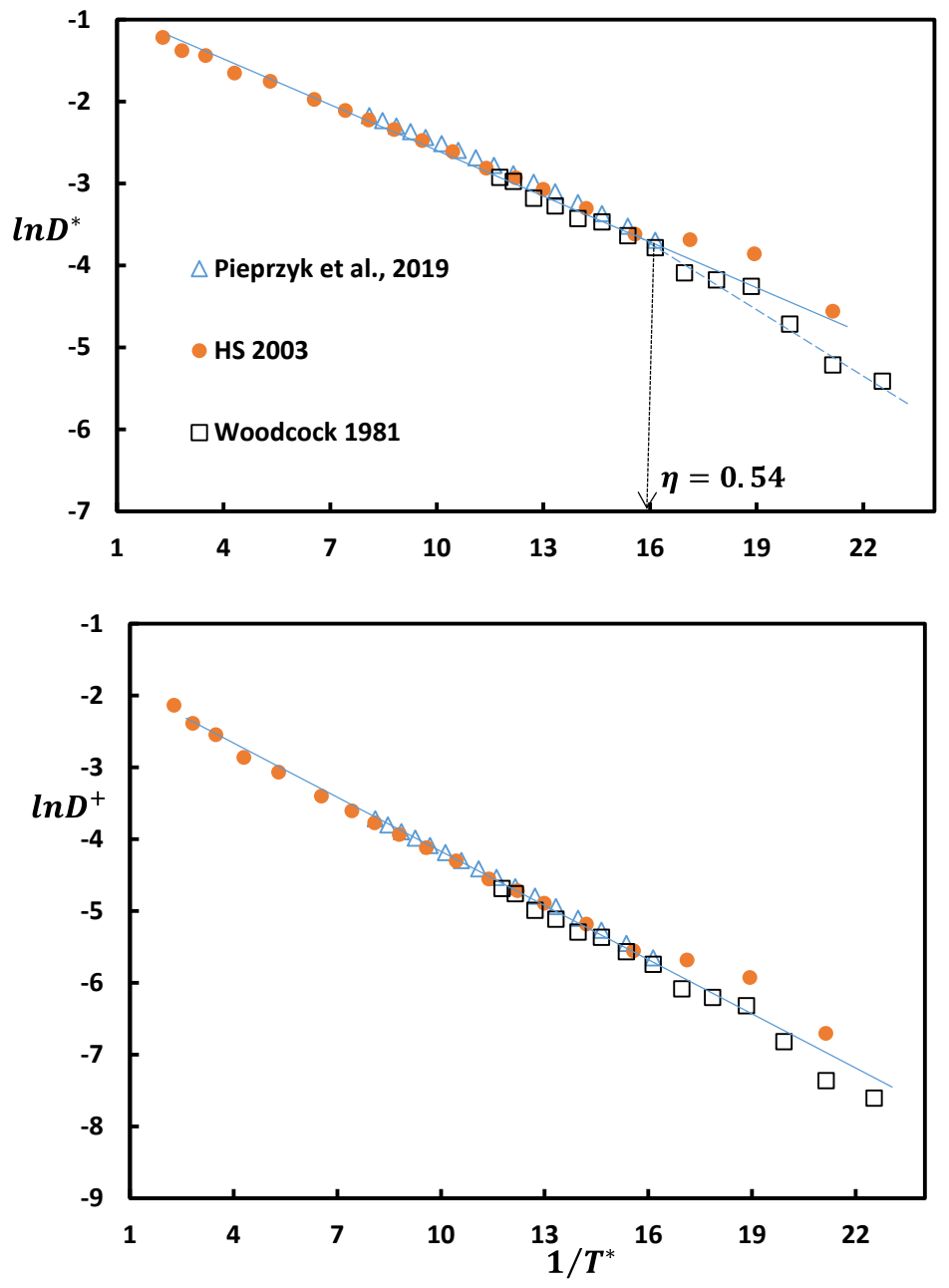

**Figure 19**. Arrhenius plots for diffusivity, above: $lnD^* \sim T^{*-1}$, below: $lnD^+ \sim T^{*-1}$. The simulation data are from Refs. [3,20,47].

Figure 18 depicts the $ln(\tau/\tau_0) \sim T^{*-1}$ and $ln\mu^* \sim T^{*-1}$ plots, respectively. It can be observed that at $\eta \approx 0.555$ ($1/T^* = 21$), the Arrhenius law fails: the slope changes between two distinct regions. This suggests that the HS system is fragile, exhibiting glass transition characterized by the relaxation time and viscosity changes at $\eta_g \approx 0.555$. Interestingly, the heat capacity is observed to peak at the same density, as discussed earlier, Figure 14. Thus, defining the glass transition point by the heat capacity peak yields a result consistent with that derived from the transport properties.

Figure 19 shows plots of reduced diffusivities: $lnD^* \sim T^{*-1}$ and $lnD^+ \sim T^{*-1}$. From $lnD^* \sim T^{*-1}$ plot, we see that the Arrhenius law fails at $\eta \approx 0.54$, in consistent with the observations from the viscosity and relaxation time (Figure 18). An interesting observation from the $lnD^+ \sim T^{*-1}$ plot is that utilizing the reduced property, $D^+$, indeed improves the correlation.

For the transport properties, another approach often used is the so-called free volume theory,

originated from an empirical correlation proposed in 1950th [51], which is effective in data correlations for liquids. Theoretical refinements have been incorporated into the model [1,2] and as a result, the free volume theory has become one of the most accepted models. However, the model has an intrinsic limitation: it adopts the van der Waals free volume expression, $v_{vf} = v - v_0$, which is primarily applicable to low-density gases, whereas the free volume model itself is intended for liquids or dense fluid [51,1,2]. Taking viscosity as an example:

$$ln\phi^* = A\exp\left(-\frac{Bv_0}{v_{vf}}\right) = A\exp\left(-\frac{Bv_0}{v - v_0}\right) \quad (44)$$

The first part of Eq.(44), $ln\phi^* \propto exp\left(-\frac{Bv_0}{v_{vf}}\right)$, represents the so-called free volume model [1], while the second part, $ln\phi^* \propto \exp\left(-\frac{Bv_0}{v-v_0}\right)$, is generally employed for data correlation [51]. In practical applications, all three constants, $A, B$ and $v_0$ are treated as adjustable parameters for data fitting. The issue arises with parameter $v_0$, which originates from the free volume. Strictly speaking, its value should come from some thermodynamic property, such as EoS. The successes of the 2nd part of Eq.(44) (with $v_0$ as a parameter from transport properties) do not necessarily validate the free volume theory (the first part of Eq. (44)). This is because the van der Waals free volume, $v - v_0$, is a poor expression for liquids or dense gases.

The present author has proposed an alternative free volume model based on a power law approach [48], which incorporates a well-defined geometric free volume. For the reduced fluidity of HS fluids, the model is expressed as:

$$\phi^* = \frac{\phi}{\phi_0} = \left(\frac{v_f}{v}\right)^{\xi} \quad (45)$$

where $v_f$ is the geometric free volume [52,53], $v = 1/\rho^*$, and $\xi$ is a constant. From computer simulations [52,53], a distribution function for the local free volume, $v_l$, is derived [48]:

$$f(v_l) = \frac{b}{v_f}\left(\frac{v_l}{v_f}\right)^{\alpha} \exp\left[-\beta\left(\frac{v_l}{v_f}\right)^{\gamma}\right] \quad (46)$$

where the parameters are determined as follows: $\alpha = 0.275$, β=3.18, $\gamma = 0.47$, $b = 6.947$ for the density range, $0.699 \le \rho^* \le 1.03$ and $v_f$ is the mean free volume at a given density. Using the same arguments for the Voronoi-cell free volume distribution function [2,54], the excess entropy can be expressed in terms of the free volume:

$$\frac{s^{ex}}{k_B} = -C_0\int_0^{\infty} f(v_l)ln[f(v_l)]\,dv_l = -C_0 ln\left(\frac{v_f}{v}\right) + C_1 \quad (47)$$

where $C_0$ and $C_1$ are constants. The thermodynamic free volume, $v_{tf}$, is defined in terms of the excess entropy (see ref.[55] and references cited therein):

$$ln\left(\frac{v_{tf}}{v}\right) = s^{ex} = -\int_0^{\rho}(Z-1)\frac{d\rho}{\rho} \quad (48)$$

Now for viscosity we have:

$$\emptyset^* = \left(\frac{v_{tf}}{v}\right)^{\kappa_v} = exp(\kappa_v s^{ex}) \quad (49)$$

Similarly, for diffusion coefficient:

$$D^* \equiv \left(\frac{v_{tf}}{v}\right)^{\kappa_d} = exp(\kappa_d s^{ex}) \quad (50)$$

where $\kappa_v$ and $\kappa_d$ are constants. Eq. (49) and Eq. (50) are particularly noteworthy as they establish a connection between transport properties and thermodynamic properties, which can be directly calculated from an EoS. In particular, $\emptyset^* = exp(\kappa_v s^{ex})$ and $D^* = exp(\kappa_d s^{ex})$ are known as the excess entropy scaling laws [48,49]. For the HS system, the only parameter to be determined from the transport property is $\kappa_v$ or $\kappa_d$, and no additional parameters are required for the free volume or the entropy.

It is interesting to note that using different EoS to calculate $v_{tf}/v$ or $s^{ex}$, results in different known models for transport properties [48]. Specifically, if the EoS from Heyes and Woodcock (1986) [56] is adopted, $Z = 1 + 4\eta/(1 - e_0\eta)^2$ ($e_0$ is a constant), the well-know Doolittle-Cohen-Turnbull (DCT) free volume model, Eq.(44), is recovered. Thus, the DCT model is formally "correct" and hence its successes in describing transport properties are not merely coincident.

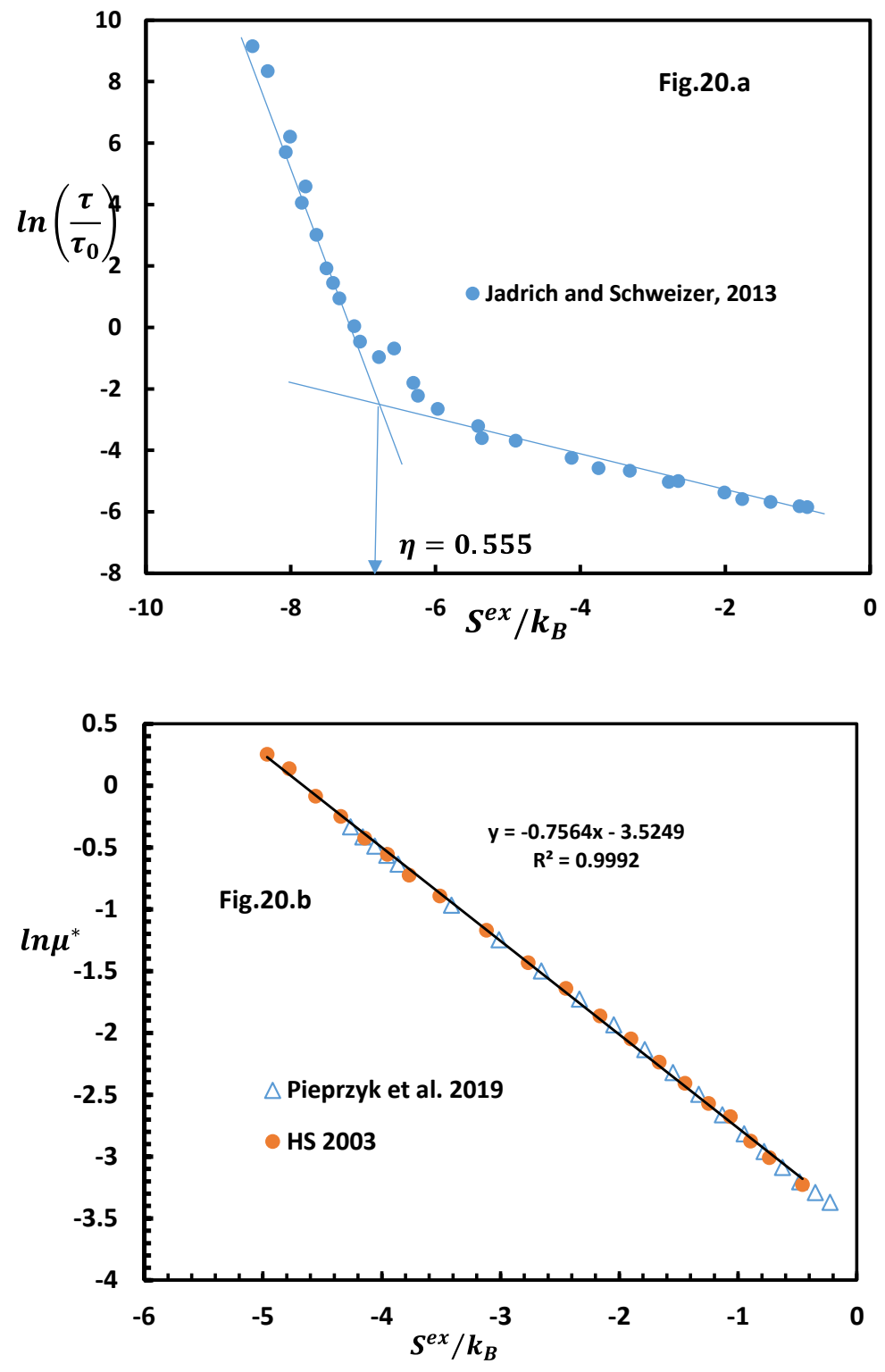

**Figure 20**. The entropy scaling laws for $\alpha$ relaxation time (a) and viscosity (b) of the HS system. Data points are from Refs.[8,20,47]. The excess entropy is calculated from Eq.(24), Eq.(28) with $\eta_J = 0.64$.

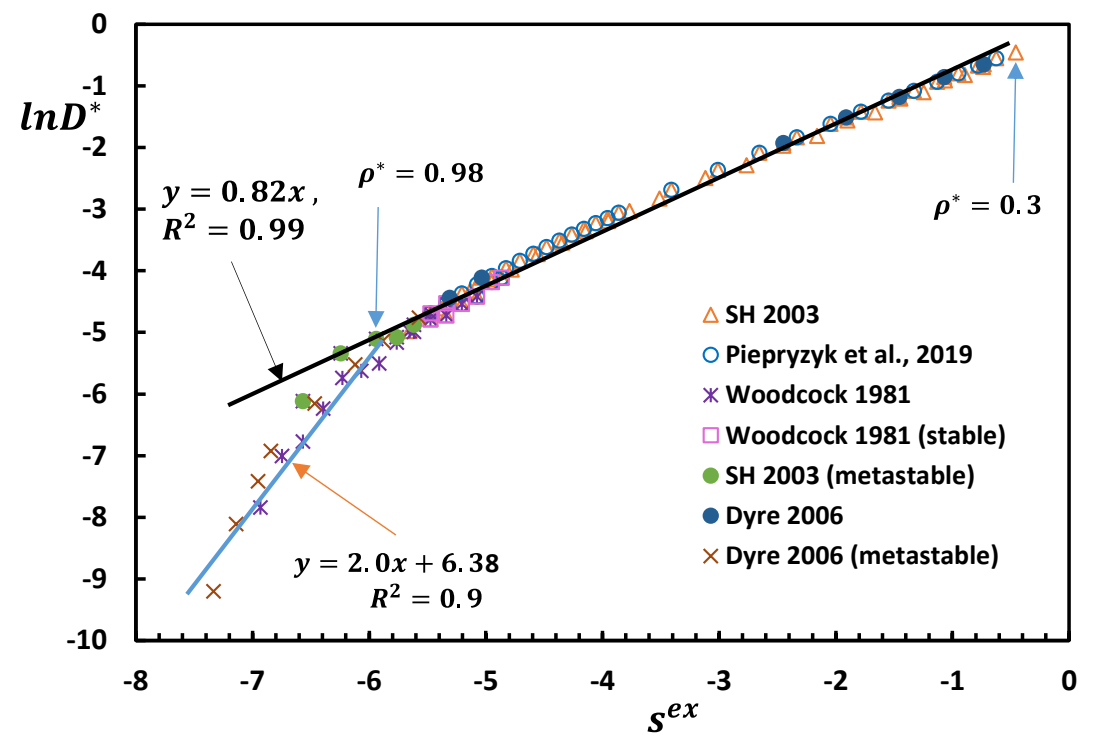

**Figure 21**. The entropy scaling law for diffusivity. The data sets are from refs.[3,20,47,57].

Figure 20 and Figure 21 illustrate the entropy scaling laws for the alpha-relaxation time, viscosity and diffusivity, respectively. The results align with those observed in the Arrhenius plots. For the relaxation time and viscosity plots, the scaling laws fail at $\eta = 0.555$: different slopes are required. With the diffusivity plot, the entropy scaling law fails at $\rho^* \approx 0.98$ ($\eta = 0.51$).

## VI. Concluding remarks

In this work, we propose a hard-sphere equation of state based on a Gamma-distribution based potential energy landscape theory. Specifically, a parameterised or jammed-state (rando jammed packing) dependent EoS, $Z(\eta_J)$, enabling calculations of the compressibility factor over the metastable and glassy region, is derived. For any given value of the jammed packing fraction, $\eta_J$, across a range from ~0.62 to ~0.66, an EoS for a path from the stable fluid to the jammed state, $\eta_J$, can be generated. This parameterized EoS provides a way to numerically "map" the entire metastable and glassy region, as projected by computer simulations [4,8].

Two specific cases have been discussed in detail. The first case studies an EoS that leads to the random close packing state, $\eta_{rcp} = 0.64$, which is generated from the parameterised EoS by setting $\eta_J = \eta_{rcp}$. The reliability of the EoS along this path is supported by its high accuracies in predicting pressure (compressibility), and various thermodynamic properties, including the isothermal compressibility [37] and excess entropy [38-40] up to a deeply metastable region. In the stable fluid region, the accuracy is comparable to the best simulation results available [20, 31,32].

Notably, the PEL based EoS enables the analysis of the system's stability by leveraging the inherent-structure components, $Z_{IS} + Z_{vib}$. For the HS system along the $\eta_{rcp}$ path, it is found that the metastable system achieves its maximum mechanical stability at $\eta = 0.573$ ($\rho^* \approx 1.1$), which is referred as the Sastry density.

The second case discussed is an ideal glass at $\eta_J = 0.661$. This state is achieved under two constraints: (1) the system passes through the stable fluid region; (2) the freezing entropy [45] is entirely converted into the configurational entropy at the Kauzmann point, $\eta_K = 0.64$. The reliability of the generated ideal glass is validated through the prediction of configurational entropy. The results presented in this work may serve as a guide for future studies aimed at exploring the behavior and properties of ideal glasses.

The final part of this work focuses on the transport properties—specifically, the alpha-relaxation time, viscosity, and diffusivity for the HS system. The Arrhenius plots suggest a correlation, $ln\phi^* \propto 1/T^*$, or for the HS case, $ln\phi^* \propto \rho^* Z$. The later exhibits behavior similar to a correlation, $ln\phi^* \propto Z$ [7]. Another approach, the excess entropy scaling law [48,49,50], is also investigated. This approach is equivalent to the (geometric) free volume power law [48]. Both the Arrhenius plots and the entropy scaling modeling break down at $\eta = 0.555$, meanwhile $Z_J + Z_{IS} + Z_{vib} > 0$, namely the inherent structures and jamming begin to emerge.

## Supplementary Materials

### The parameterised EoS (summary)

$$Z = 1 + Z'_v + \frac{b_0\eta}{1-\eta/\eta_{cp}} + \frac{b_1\eta^{n_1}}{1-\eta/\eta_J} + b_2\eta^{n_2} + b_3\eta^{n_3} \quad (S1)$$

$$Z'_v = \sum_{n=2}^{11}\left(B_n - b_0\alpha_{cp}^{n-1}\right)\eta^{n-1} \quad (S2)$$

where $B_2 = 4$, $B_3 = 10$ and

$$B_n = -1.8752 + 1.2872n + 0.944n^2, n \geq 4 \quad (S3)$$

$$b_i(\eta_J) = \sum_{k=0}^{3} c_{i,k}\,\eta_J^k, \qquad i = 1,2,3 \quad (S4)$$

The constants are listed in Table 3 and Table 4 (main text).

Table S1 lists comparison between the simulation results [20] and calculations for the compressibility in the stable fluid region. The differences appear only after the 4th digits in the fractional part.

Table S4 and Figure S1 list the values of parameters from which Table S2, Eq.(S4), is obtained.

**Table S1**. Compressibility data from Computer simulations [20] vs EoS calculations at selected densities.

| $\rho^*$ | $\eta$ | Z (simulation) | Z (calculation) |
|---|---|---|---|
| 0.05 | 0.02618 | 1.111917 | 1.111917 |
| 0.1 | 0.05236 | 1.239720 | 1.239721 |
| 0.2 | 0.10472 | 1.553604 | 1.553615 |
| 0.3 | 0.15708 | 1.968231 | 1.968243 |
| 0.4 | 0.209439 | 2.521620 | 2.521629 |
| 0.5 | 0.261799 | 3.269404 | 3.269398 |
| 0.6 | 0.314159 | 4.294977 | 4.294964 |
| 0.7 | 0.366519 | 5.726960 | 5.726989 |
| 0.8 | 0.418879 | 7.770090 | 7.770092 |
| 0.9 | 0.471239 | 10.762722 | 10.762733 |
| 0.928 | 0.485899 | 11.843840 | 11.843831 |
| 0.938 | 0.491135 | 12.262842 | 12.262819 |

**Table S2**. The EoS parameters, Eq.(24), for the path $\eta_J = 0.64$

| $b_1$ | $b_2$ | $b_3$ | $AAD\%$ |
|---|---|---|---|
| 1.44910E+03 | -1.97379E+04 | 6.28379E+08 | 0.00065 |

**Table S3**. Values of the parameters of Eq.(24) for the path leading to the ideal glass

| $\eta_{igj}$ | $b_0$ | $n_1$ | $n_2$ | $n_3$ | $b_1$ | $b_2$ | $b_3$ | $AAD\%$ |
|---|---|---|---|---|---|---|---|---|
| 0.661 | 10.0 | 14 | 14 | 42 | 261.7775 | -3.170158E+03 | -6.52744E+08 | 0.0015 |

**Table S4** Values of fitting parameters and AAD.

| $\eta_J$ | $\boldsymbol{b_1}$ | $\boldsymbol{b_2}$ | $\boldsymbol{b_3}$ | $\boldsymbol{AAD\%}$ |
|---|---|---|---|---|
| 0.62 | 1137.846 | -18089.361 | 4.62985E+08 | 0.00065 |
| 0.625 | 1208.983 | -18472.311 | 5.15637E+08 | 0.00065 |
| 0.63 | 1284.571 | -18874.725 | 5.59709E+08 | 0.00065 |
| 0.635 | 1364.612 | -19296.602 | 5.96767E+08 | 0.00065 |
| 0.64 | 1449.105 | -19737.942 | 6.28379E+08 | 0.00065 |
| 0.645 | 1538.049 | -20198.746 | 6.56110E+08 | 0.00065 |
| 0.65 | 1631.446 | -20679.012 | 6.81529E+08 | 0.00065 |
| 0.655 | 1729.295 | -21178.742 | 7.06201E+08 | 0.00065 |
| 0.66 | 1831.595 | -21697.936 | 7.31693E+08 | 0.00065 |

AAD% the absolute average deviation for the stable liquid region.

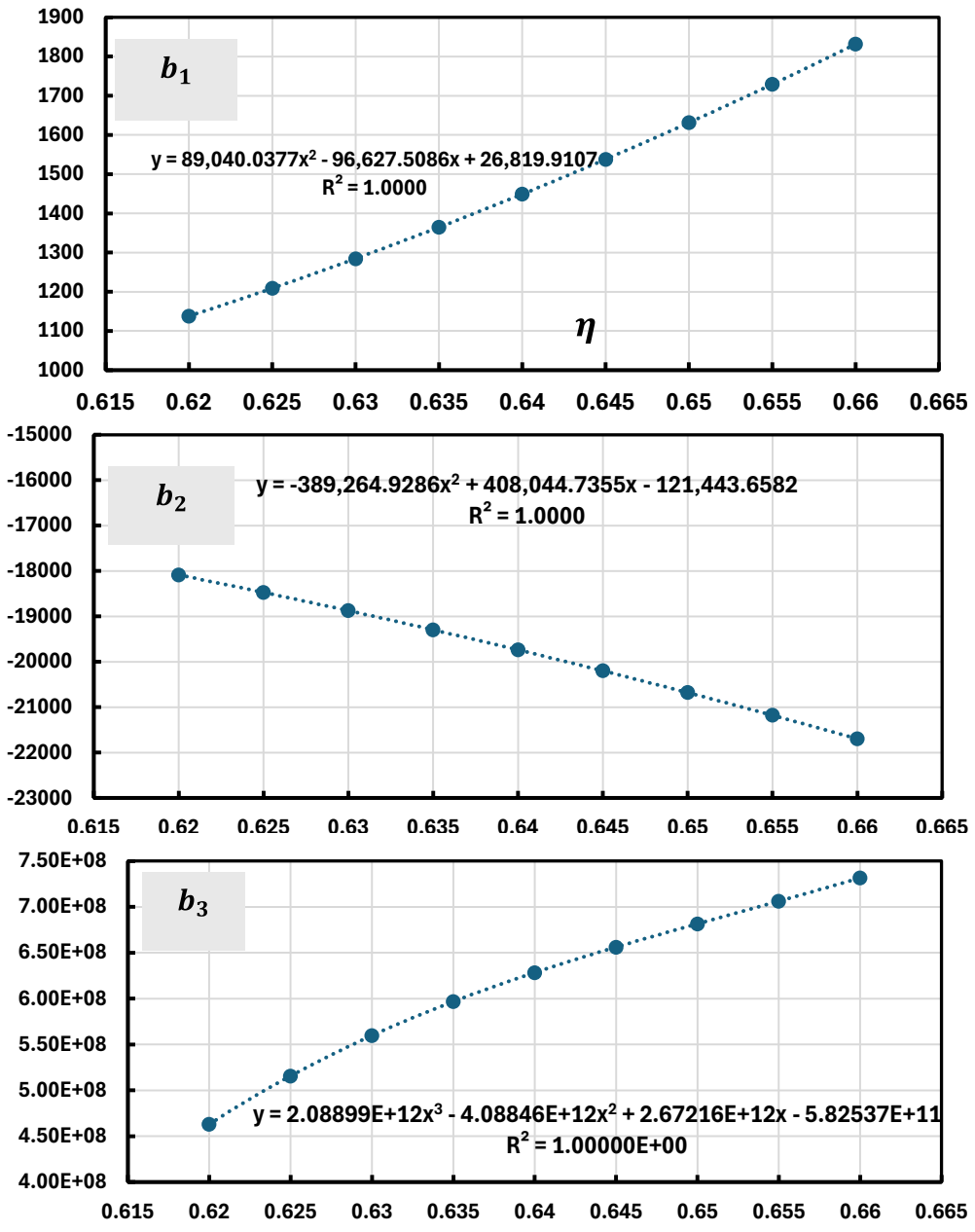

**Figure S1**. correlations of parameters $b_1$, $b_2$ and $b_3$ with packing fraction.

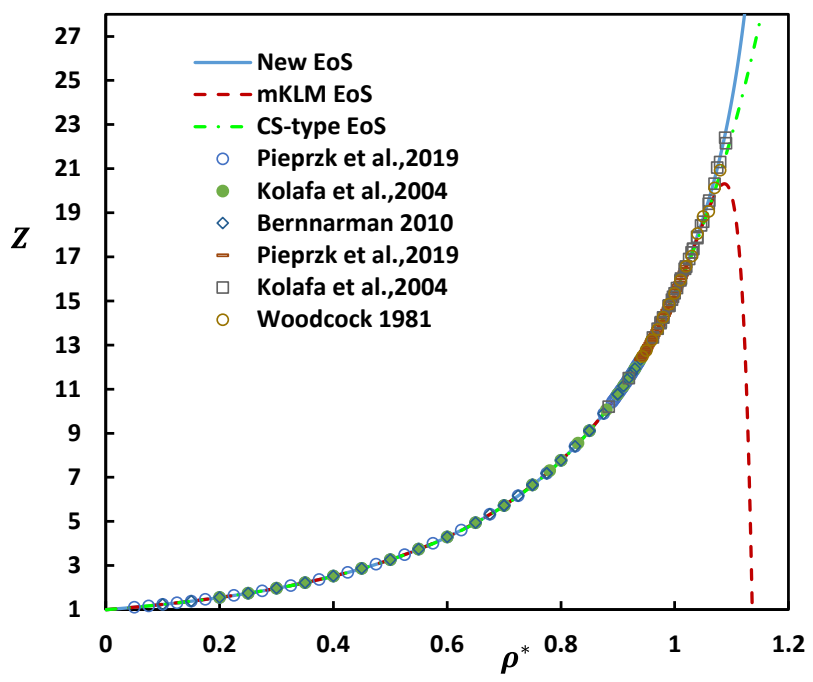

**Figure S2**. A phase diagram in $Z\sim\rho^*$ space, calculations vs simulations. The mKLM EoS reaches a maximum at $\rho^* = 1.087$ and falls to zero at $\rho^* = 1.137$.

**Table S5.** Compressibility in the metastable region

| data author | $\rho^* = 1$ | | $\rho^* = 1.05$ | | $\rho^* = 1.08$ | | $\rho^* = 1.1$ | | $\rho^* = 1.15$ | |
|---|---|---|---|---|---|---|---|---|---|---|
| | Z | RD % | Z | RD % | Z | RD % | Z | RD % | Z | RD% |
| Woodcock, 1981 [3] | 15.34 | -0.17 | 18.83 | 0.14 | 20.93 | -2.68 | | | | |
| Rinto&Torquato,1996 [33] | 15.21 | -1.02 | 19.00 | 1.04 | 21.78 | 1.29 | 24.27 | 0.27 | | |
| Speedy 1997 [34] | 15.59 | 1.46 | 18.70 | -0.55 | 21.48 | -0.12 | | | | |
| Kolafa et al.,2004 [32] | 15.35 | -0.10 | | | | | | | | |
| Wu&Sadus, 2005 [35] | 15.35 | -0.10 | 18.47 | -1.77 | 20.84 | -3.09 | 23.60 | -2.50 | 34.82 | -1.00 |
| Kolafa, 2006 [19] | 15.46 | 0.61 | 18.59 | -1.14 | 21.30 | -0.96 | | | | |
| Bannerman et al.,2010 [31] | 15.32 | -0.30 | | | | | | | | |
| Hermes&Dijkstra, 2010 [36] | 15.24 | -0.82 | 19.21 | 2.16 | 22.43 | 4.30 | 24.93 | 2.99 | 37.67 | 7.11 |
| Pieprzyk et al., 2019 [20] | 15.35 | -0.10 | 18.59 | -1.14 | | | | | | |
| Zhou &Schweizer, 2020 [37] | 15.45 | 0.55 | 19.04 | 1.26 | 21.78 | 1.27 | 24.02 | -0.76 | 33.02 | -6.11 |
| **Algebraic mean** | **15.37** | | **18.80** | | **21.51** | | **24.21** | | **35.17** | |
| New EoS | 15.35 | 0.001 | 18.80 | 0.25 | 21.68 | 1.12 | 24.28 | 0.81 | 37.00 | 5.27 |
| CS-type EoS [30] | 15.30 | -0.33 | 18.45 | -1.61 | 20.73 | -3.31 | 22.45 | -6.79 | 27.60 | -21.5 |

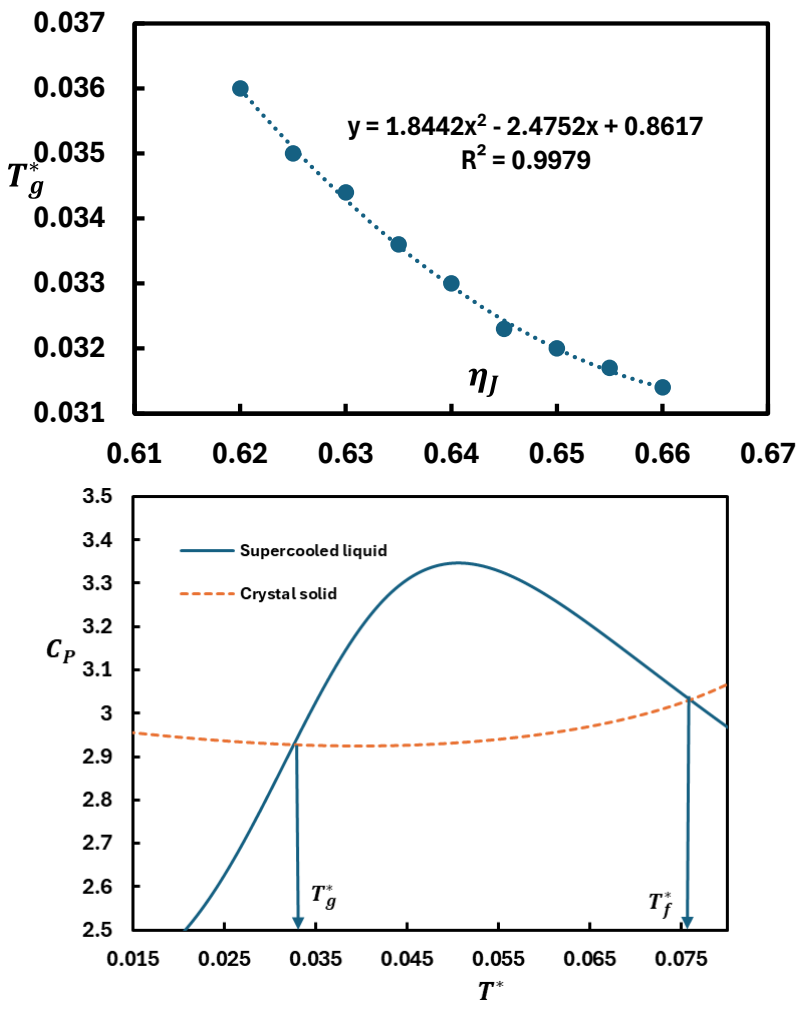

**Figure S3**. Glass transition temperature vs $\eta_J$ (above) and determination (below).

Figure S3 shows that the glass transition temperature (density) is also dependent on the jammed density (packing fraction).

## EoS and heat capacity for the crystalline solid

The EoS for the hard sphere solid is from Ref.[20]:

$$Z_s = \frac{3}{1-w} + \frac{1}{w+1.5} + Ae^{B(1-w)} + Ce^{D(1-w)} + E \quad (S5)$$

where $w = \rho^*/\rho_{cp}^*$, $\rho_{cp}^* = 1.4142$. $A = 0.025882$, $B = 8.689$, $C = 3.5433 \times 10^{-6}$, $D =$

$34.377$ and $E = -0.85973$. The heat capacity of the solid is given by Eq.(26) (main text).

Now the temperature ($T_s^* = 1/(\rho^* Z_s)$ of the heat capacity is fitted with a polynomial function:

$$C_{p,s}^{ex} = aT_s^{*6} + bT_s^{*5} + cT_s^{*4} + dT_s^{*3} + eT_s^{*2} + fT_s^* + g \quad (S6)$$

The fitting constants are listed in Table S6.

Eq.(S6) are employed for determining the heat capacity in Eq.(33), Eq.(34) etc., $C_p(crystal) = C_{p,s}^{ex}$, where temperature is replaced with the same value as used for the liquid phase, $T_s^* = T^*$. The integration can be carried out numerically.

**Table S6.** constants for heat capacity of solid fitting, Eq.(S6)

| $a$ | $b$ | $c$ | $d$ | $e$ | $f$ | $g$ |
|---|---|---|---|---|---|---|
| 2.79338E+06 | -5.33610E+05 | 4.23905E+04 | -1563.51 | 70.43419 | -3.81075 | 3 |

The AAD of the Eq.(S3) is 0.0034% against the simulation values [20].